# Holocrine Secretion and Kino Flow in Angiosperms: Their Role and Physiological Advantages in Plant Defence Mechanisms

Paulo Cabrita

Dr. Paulo Cabrita

Soderstr. 39, 64287 Darmstadt, Germany

**ORCID ID**: 0000-0002-2620-3573

**Corresponding author E-mail**: paulocabrita@hotmail.com

**Current Address**: Weserstr.36, 37081 Göttingen, Germany

**Telephone**: +49 551 48942338

## Acknowledgements

This research did not receive any specific grant from funding agencies in the public, commercial, or not-for-profit sectors. The author would like to thank to two anonymous reviewers for the careful reading and the very helpful suggestions and criticism of this manuscript.

This is a post-peer-review, pre-copyedit version of an article published in *Trees*. The final authenticated version is available online at: https://doi.org/10.1007/s00468-020-01990-z

## Abstract

Kino is a plant exudate, rich in polyphenols, produced by several angiosperms in reaction to injury of the cambium. It flows out of kino veins, which compose an anatomically distinct continuous system of tangentially anastomosing canals produced by the cambium upon damage, encircling plant stems and branches. Kino is loaded into the vein lumen by autolysis of a cambiform epithelium lined by suberized cells that separate kino veins from the surrounding axial parenchyma. A model describing kino flow is presented to investigate how vein distribution and structure, as well as the loading, solidification, and viscosity of kino affect flow. Considering vein anatomy, viscosity, and a time-dependent loading of kino, the unsteady Stokes equation was applied. Qualitatively, kino flow is similar to resin flow observed on conifers. There is an increase in flow towards the vein open end, with both pressure and flow depending on the vein dimensions, properties, and loading of kino. However, kino veins present a much smaller specific resistance to flow compared to resin ducts. Also, unlike resin loading in conifers, the loading of kino is not pressure-driven. The pressure and pressure gradient required to drive an equally fast flow are smaller than what is observed on the resin ducts of conifers. These results agree with previous observations on some angiosperms and suggest that flow within internal secretory systems derived from autolysing epithelia may have lower metabolic energy costs; thus presenting physiological advantages and possibly constituting an evolutionary step of angiosperms in using internal secretory systems in plant defence mechanisms compared to resin flow in conifers. Understanding of how these physiological and morphological parameters affect kino flow might be useful for selecting species and developing more sustainable and economically viable methods of tapping gum and gum resin in angiosperms.

# 1. Introduction

Gum and gum resin ducts of many angiosperms share several anatomical and physiological similarities with resin ducts (Evert, 2006; Nair, 1995). In both internal secretory systems the loading of the duct lumen can be granulocrinous, i.e. through the fusion of Golgi or endoplasmic reticulum vesicles with the plasmalemma, releasing their contents through the cell wall, or eccrinous involving the passage of small molecules directly through the plasmalemma and cell wall (Beck, 2010; Evert, 2006). In some cases, however, the loading of the duct lumen occurs through the lysis of epithelium surrounding it, in what is termed holocrine secretion (Nair, 1995). Considering the role that gum plays in plant defence mechanisms and the similarities between gum-producing angiosperms and resin-producing species, the unsteady Stokes equation can be similarly applied to describe flow inside gum and gum resin ducts (Cabrita, 2018). The unsteady Stokes equation describes what is called Stokes flow or creeping flow, where the viscous forces dominate the entire flow over the advective inertial forces (Shames, 2003). This is a typical situation in flows where the velocities are very slow, the viscosities are very large, or the length-scales are very small. In plant secretory systems, very slow flows are observed within systems with very small length-scales, e.g. resin ducts (Cabrita, 2018; Schopmeyer *et al*, 1954).

This study aims to investigate the physiological and anatomical advantages of holocrine secretion on the flow within internal secretory systems of angiosperms by applying the unsteady Stokes equation, using kino veins produced by the cambium as model. A comparison will be made with the pressure-driven granulocrine loading of resin (Cabrita, 2018). The main focus is on the anatomical and physiological aspects of kino veins, as well as exudate properties, and how these affect flow and contribute to plant defence strategies. The author hopes that this study contributes to a better understanding of the physiological processes behind holocrine secretion and flow of plant exudates. Additionally, the results presented could help the development of better tapping methods envisaging a more sustainable and economically viable extraction of plant exudates. Before introducing the mathematical model and to best understand and interpret it, the state of the art in the development, structure, and physiology of kino veins is presented.

## 1.1. What are kino veins?

A typical example of internal secretory systems of plants where lysis of the epithelium occurs are the kino veins found in the secondary vascular system of some angiosperms (Table 1). Kino veins form an anatomically distinct continuous system of tangentially anastomosing lacunae or canals occurring as annular zones, usually incomplete, running longitudinally, often encircling the plant stem and branches, and embedded in parenchyma that developed in response to trauma of the cambium, termed traumatic parenchyma (Hillis, 1987; Subrahmanyam and Shah, 1988). When ruptured, they release kino[1], a viscous polyphenolic exudate. Their distinct colouring, reflecting kino composition, gives kino lacunae or canals the appearance of veins running through the wood, making them an easily distinct feature (Angyalossy *et al*., 2016). Kino veins are usually larger than other plant internal secretory systems and their development and physiology is best known from the genera *Acacia*, *Azadirachta*, *Butea*, *Corymbia, Eucalyptus*, *Prosopis,* and *Pterocarpus*. Similarly to many gum and resin ducts, kino veins arise from injury of the cambium in response to specific stress. However, the extension, distribution, and ability to produce kino veins seem to be species- and trauma-dependent (Eyles and Mohammed, 2003; Hillis, 1987; Setia, 1984b; Shearer

---

[1] The exact origin of the word kino remains yet unresolved. Some authors claim it is of West African origin, from the word *kāno* or *keno*, meaning gum in Mandika language, to name the exudate extracted from *Pterocarpus erinaceus* Poir, used by native populations for medicinal purposes, and first introduced commercially into European medicine as an astringent in 1757 by John Fothergill. Others believe that it comes from the Indian word *kini* or *kuenee* to describe the astringent exudation obtained from *Butea monosperma* (Lam.) Taub. (Maiden, 1901; McGookin and Heilbron, 1926).

**Table 1** – Plant genera known to yield kino following the APG IV system (APG IV, 2016; Jacobs, 1937; Garratt, 1933; Groom, 1926; Hegnauer, 1994; Kühn and Kubitzki, 1993; Lambert *et al.*, 2007, 2013; Maiden, 1901; McGookin and Heilbron, 1926; Rajput *et al.*, 2005; Subrahmanyam, 1981; van Wyk *et al.*, 1983; Venkaiah, 1982; Wehmer, 1935)

| Family | Genera |
|---|---|
| Anacardiaceae | *Schinopsis* |
| Casuarinaceae | *Casuarina* |
| Cunoniaceae | *Ceratopetalum*, *Schizomeria* |
| Euphorbiaceae | *Baloghia*, *Croton*, *Jatropha*, *Macaranga* |
| Leguminosae | *Acacia*, *Berlinia*, *Brachystegia*, *Butea*, *Centrolobium*, *Dalbergia*, *Derris*, *Dipteryx*, *Lonchocarpus*, *Machaerium*, *Mezoneurum*, *Millettia*, *Parkia*, *Prosopis*, *Pterocarpus*, *Sesbania*, *Spatholobus*, *Xylia* |
| Malvaceae | *Adansonia*, *Bombax*, *Kydia* |
| Meliaceae | *Azadirachta*, *Carapa*, *Cedrela*, *Khaya*, *Lovoa*, *Melia*, *Sandoricum*, *Swietenia*, *Toona*, *Xylocarpus* |
| Moringaceae | *Moringa* |
| Myricaceae | *Myrica* |
| Myristicaceae | All genera |
| Myrtaceae | *Angophora*, *Arillastrum*, *Corymbia*, *Eucalyptus*, *Eugenia*, *Syzygium* |
| Polygonaceae | *Coccoloba* |
| Rhizophoraceae | *Kandelia*, *Rhizophora* |
| Rubiaceae | *Uncaria* |
| Zygophyllaceae | *Guaiacum* |

*et al*. 1987; Subrahmanyam, 1981; Subrahmanyam and Shah, 1988; Tippett *et al*., 1985). Despite the consensus about the traumatic origin of kino veins, their distribution among plants is not totally known (Table 1). The last comprehensive account of kino-producing species was presented by McGookin and Heilbron (1926). However, a more complete survey could not be located. Among the several contributing factors, there is some confusion in the literature regarding the terminology of internal secretory systems of plants. For example, it is quite common to find the terms gum veins or ducts and gum used indiscriminately to refer to kino veins and kino (e.g. Greenwood and Morey, 1979; Nair *et al*, 1983; Rajput *et al*, 2005; Setia, 1984a, b). Usually, the description of the anatomy and physiology of internal secretory structures does not involve the chemical composition of the exudates. Hence, the evaluation if particular internal secretory structures are homologous or partly homologous with kino veins becomes somewhat difficult when reviewing related studies.

## 1.2. The development of kino veins

The plant hormone ethylene is thought of playing a major role in giving stimulus for the formation of kino veins after injury (Eyles and Mohammed, 2003; Greenwood and Morey, 1979; Nair *et al*., 1985; Nelson and Hillis, 1978; Tippett, 1986; Wilkes *et al*., 1989), similar to what is observed on species that produce gum (Babu *et al*., 1987; Gedalovich and Fahn, 1985a, b; Nair *et al*, 1981; Rajput and Kothari, 2005). After stimulus, periclinal and transverse divisions of a set of cambial parenchyma cells, called rosette (Prado and Demarco, 2018), result in layers of generally isodiametrically or tangentially elongated thin-walled traumatic parenchyma cells accumulating polyphenols in their vacuoles (Babu and Shah, 1987; Eyles and Mohammed, 2002; Rajput *et al*., 2005; Setia, 1984b; Skene, 1965; Subrahmanyam and Shah, 1988; Tippett, 1986). Although a xylem origin is observed, e.g. *Azadirachta indica* A.Juss. (Rajput *et al*., 2005), *Bombax ceiba* L. (Babu, 1985; Babu and Shah, 1987), *Angophora*, *Corymbia*, *Eucalyptus* (Jacobs, 1937; Tippett, 1986), *Lovoa trichilioides* Harms, *Xylocarpus gangeticus* (Prain) C.E.Parkinson (Groom, 1926), traumatic parenchyma can also originate from differentiating pre-existing phloem parenchyma, e.g. *Acacia nilotica* (L.) Delile (Subrahmanyam, 1981), *Eucalyptus* (Day, 1959; Eyles and Mohammed, 2002, 2003), *Moringa oleifera* Lam. (Subrahmanyam and Shah, 1988), *Prosopis cineraria* (L.) Druce (Venkaiah, l986). Additionally, origin in both tissues has also been reported, e.g. *Prosopis glandulosa* Torr. (Greenwood and Morey, 1979). Nevertheless, the development of kino veins is similar among kino-producing species regardless of whether is initiated in the xylem or the phloem, e.g. *Acacia nilotica* (L.) Delile (Subrahmanyam, 1981), *Azadirachta indica* A.Juss. (Rajput *et al.*, 2005; Nair *et al*., 1983), *Bombax ceiba* L. (Babu and Shah, 1987), *Eucalyptus nitens* (H.Deane & Maiden) Maiden (Eyles and Mohammed, 2002, 2003), *Kydia calycina* Roxb. (Venkaiah, 1982), *Moringa oleifera* Lam. (Subrahmanyam, 1981; Subrahmanyam and Shah, 1988), *Pterocarpus marsupium* Roxb. (Setia, 1984b), *Prosopis* (Greenwood and Morey, 1979; Venkaiah, 1986). Hence, from what is known, the development of kino veins can be summarized as follows.

After expanding up to four times their original size, the central traumatic parenchyma cells of the rosette autolyse and collapse through mechanical rupture of the walls forming a lacuna that is filled with the cellular contents rich in polyphenols thus released (Figs. 1A, B). Generally, the collapse of the rosette takes place tangentially, which makes kino veins wider, but it can also occur radially (Day, 1959). As the lacuna is formed, the traumatic parenchyma cells around its edge divide actively periclinally and anticlinally forming a peripheral cambium. The derivatives of the peripheral cambium originated inwards increase similarly in size differentiating into epithelial cells (Figs. 1B) with a dense organelle-rich cytoplasm, a large nucleus, numerous small and large vacuoles accumulating polyphenols, and a thicker and swollen lignified wall facing the vein lumen that is further increased when these cells autolyse eventually (Fahn, 1988b; Subrahmanyam and Shah, 1988; Nair, 1995; Setia 1984b). This process extends vertically, upwards, and downwards from the point of stimulus (trauma), breaking down large areas of traumatic parenchyma, and forming a tangentially anastomosing network of veins intersected by rays that are not traversed by them. The peripheral cambium also produces layers of compact cells outwards that can extend differently from one up to seven cells thick. Later in their development, the outer layer of these cells becomes suberized with thick walls forming a typical periderm lining the epithelium and separating the veins from the surrounding parenchyma tissue (Figs. 1C, D). By then, the vein lumen is lined by a layer of two to three cells thick without further development (Eyles and Mohammed, 2002; Greenwood and Morey, 1979; Hillis, 1987; Tippett, 1986). The traumatic parenchyma surrounding the kino veins eventually becomes thick-walled and lignified forming tangential series of radial and transverse bridges (Fig. 1E), occasionally with phloem fibres (Eyles and Mohammed, 2002; Subrahmanyam and Shah, 1988; Tippett, 1986). The parenchyma bridges are widest closest to the cambium, and may develop into xylem, phloem, and associated parenchyma progressively (Rajput *et al*., 2005). The development of kino veins and subsequent release of kino can occur as fast as 1-2 weeks after injury and continue from a few days to several weeks or even months (Das, 2014; Eyles, 2003; Eyles and Mohammed, 2002; Skene, 1965). Under prolonged stimulus, there is a continuous activity of the cambium in producing new kino veins, as well as maintaining active a permanent meristematic peripheral cambium and a holocrine secretion system in existing ones. Therefore, kino veins secretory activity depends on the continued meristematic activity of the peripheral cambium, as it involves the destruction of the secretory cells (epithelium).

### 1.3. Size and distribution of kino veins

Kino veins can differ considerably in size of a few centimetres to 5 m or more in length, from 1 to 7 mm in diameter, and from a few isolated veins to a dense anastomosing network of veins separated by small parenchyma bridges extending vertically for considerable distances in the stem (Fig. 2) (Dowden and Foster, 1973; Fahn, 1990; Greenwood and Morey, 1979; Loewus and Runeckles, 1977; Skene, 1965; Tippett, 1986). Occasionally, large isolated veins of a few centimetres wide and up a metre or more in length, called kino pockets (Fig. 2), may develop without parenchyma bridges, and are usually found in knobs and external swellings in the stem or even burst its limiting periderm and extend through a large area of the rhytidome (Chattaway, 1953; Day, 1959; Hillis, 1962, 1987; Phillips, 1933).

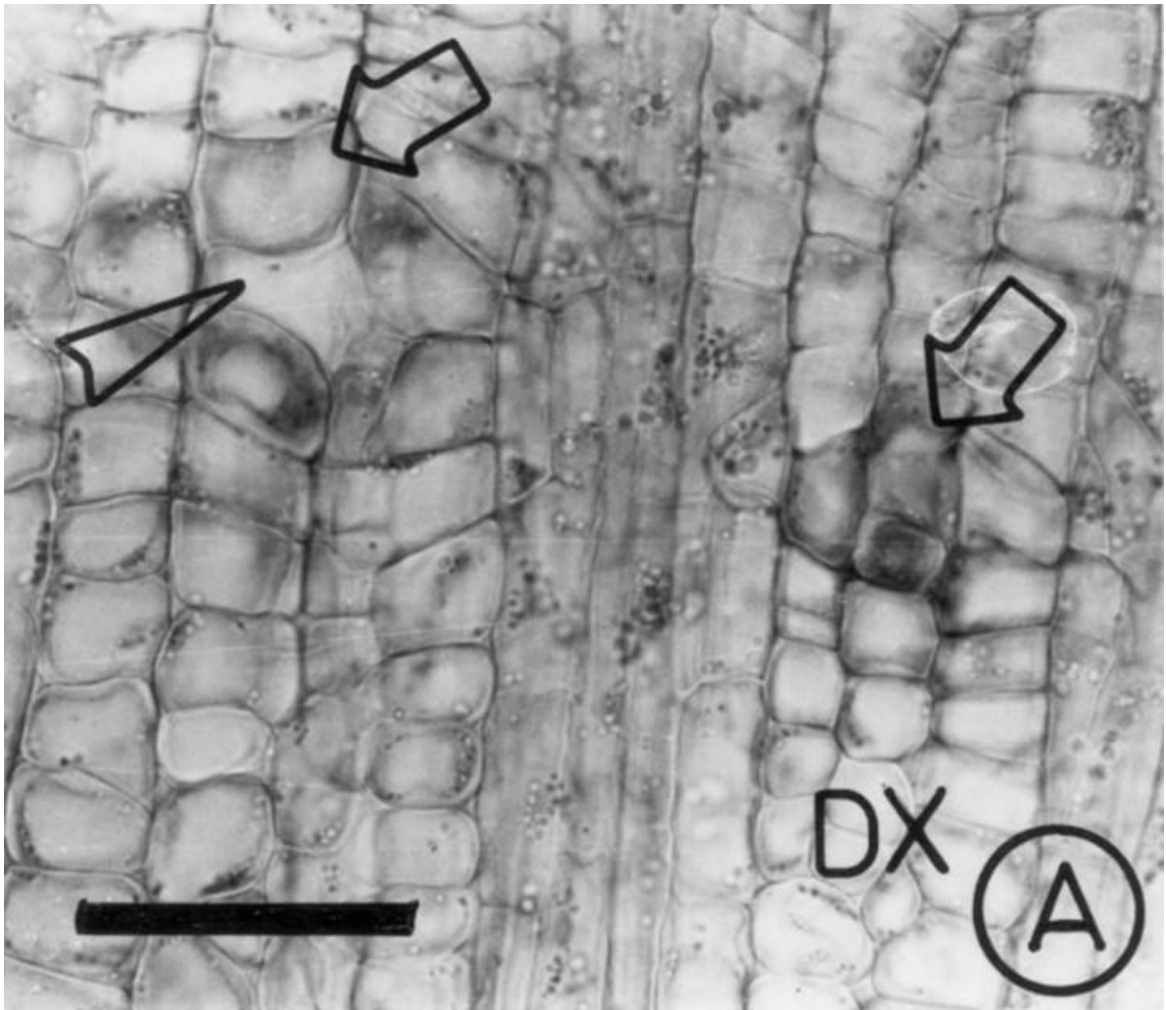

**Figure 1**

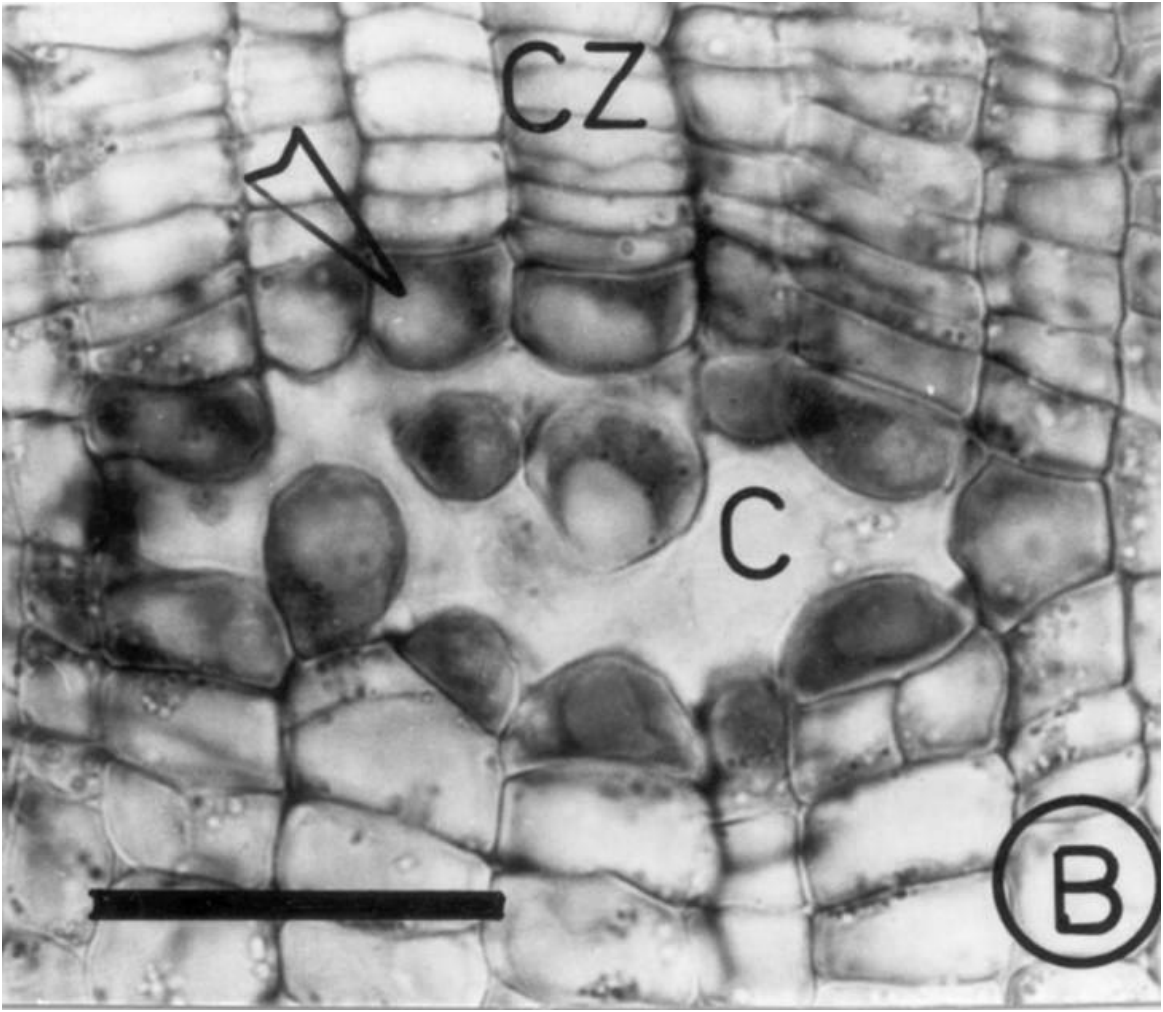

**Figure 1**

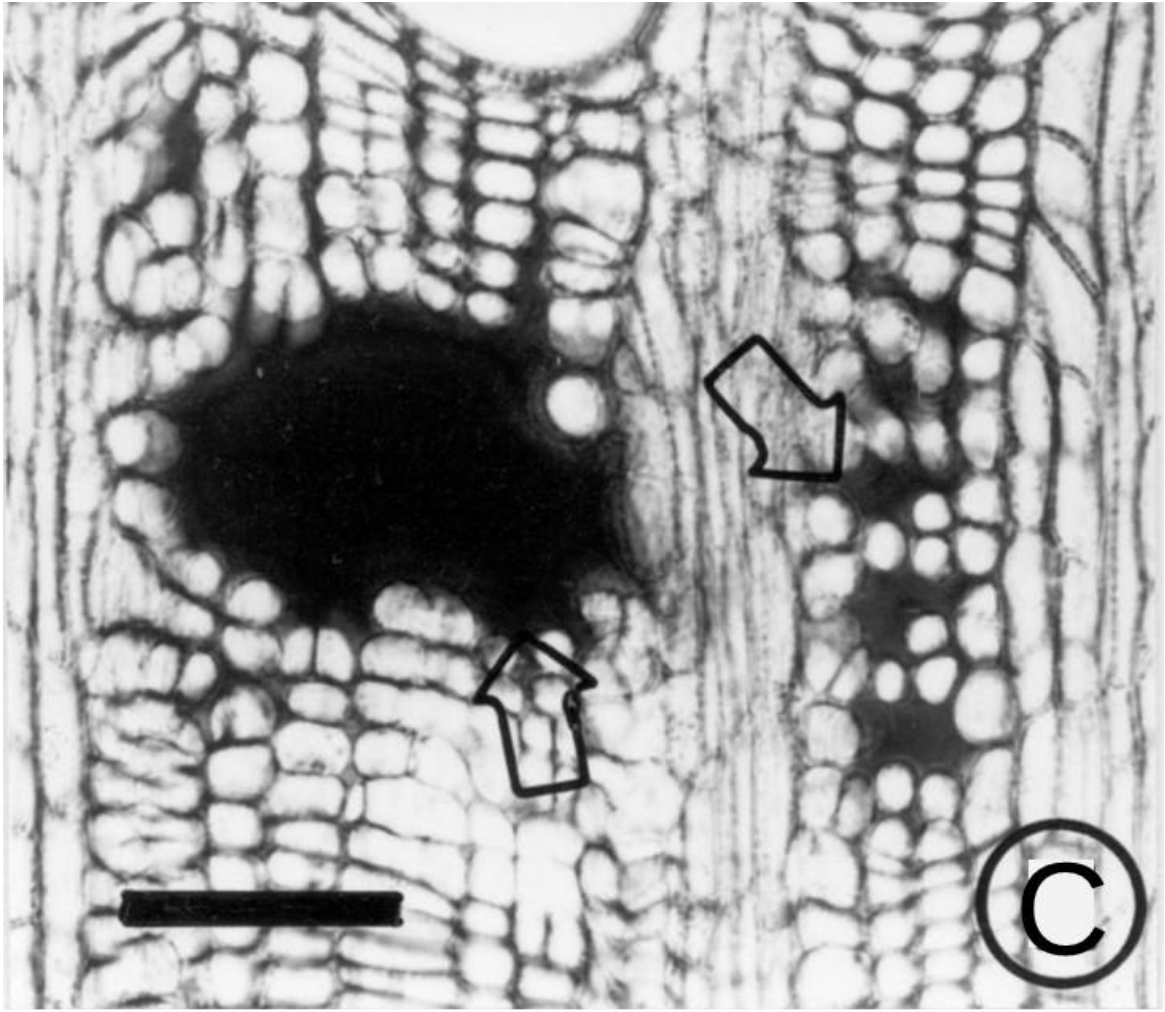

**Figure 1**

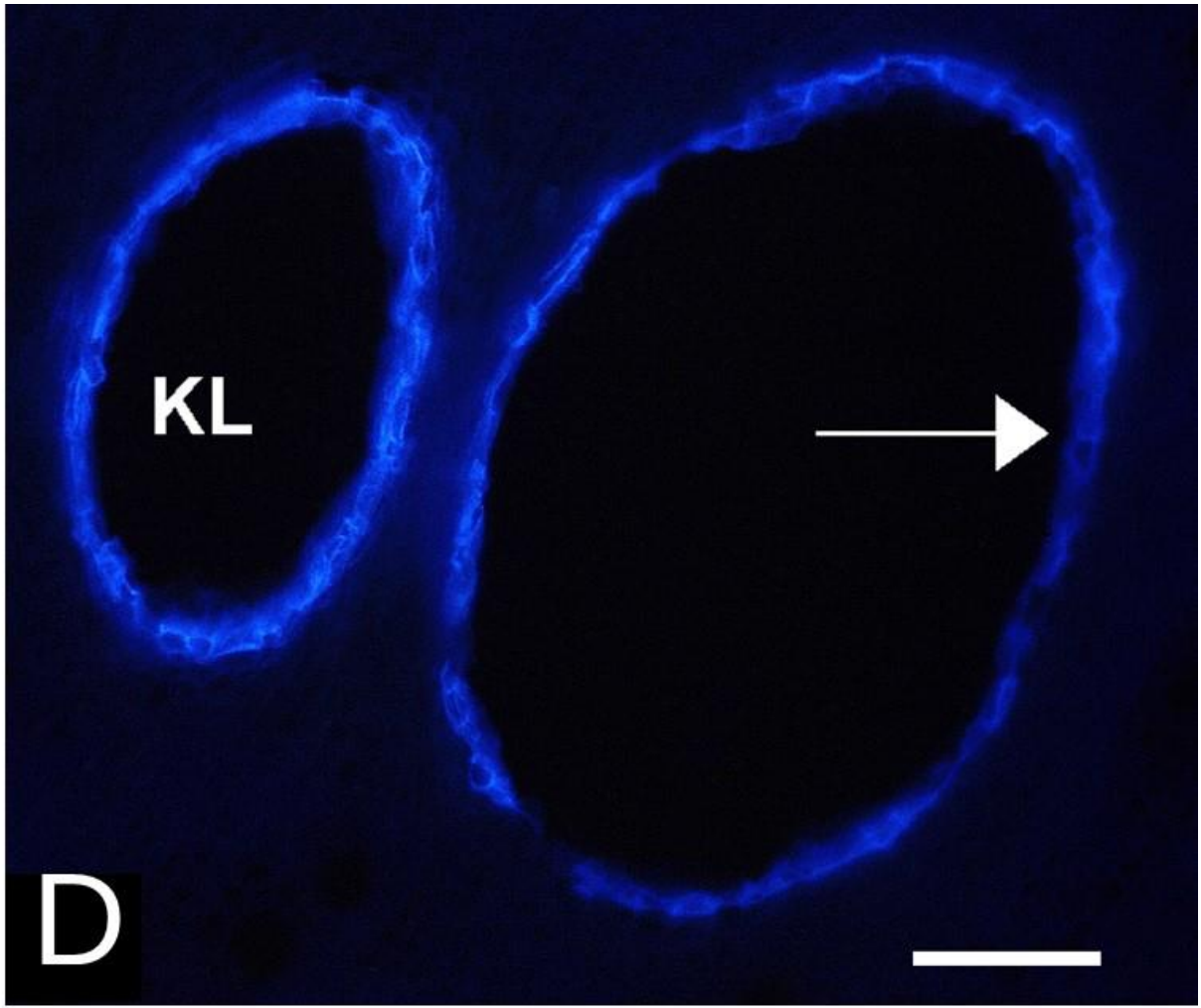

**Figure 1**

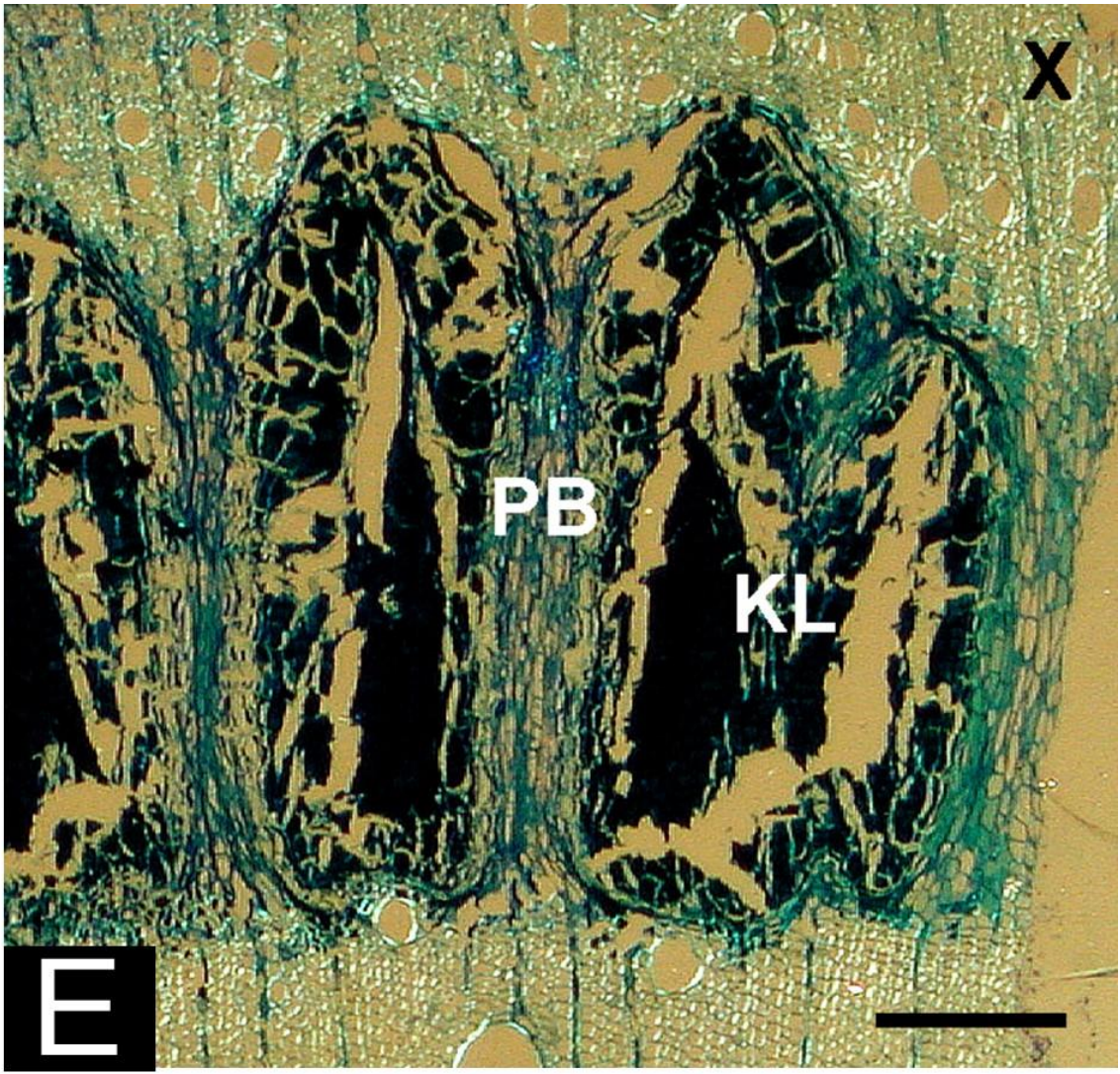

**Figure 1**

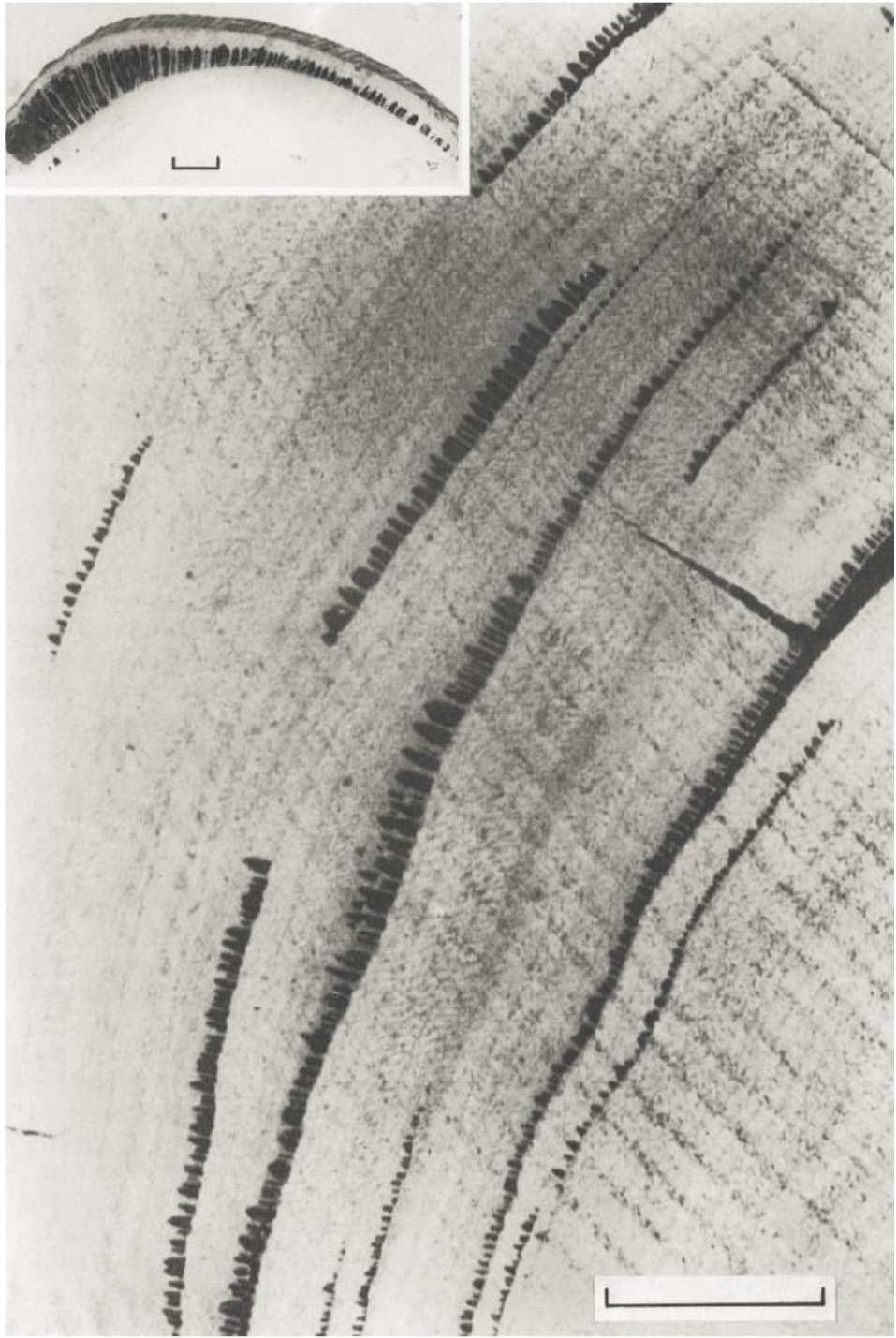

**Figure 2**

### 1.4. Kino composition, synthesis, and loading into kino veins

The kinos most valued economically come from *Acacia nilotica* (L.) Delile, *Acacia senegal* (L.) Willd., *Acacia seyal* Delile, *Butea monosperma* (Lam.) Taub.*, Coccoloba uvifera* (L.) L., *Moringa oleifera* Lam., *Prosopis juliflora* (Sw.) DC., *Pterocarpus erinaceus* Poir., *Pterocarpus marsupium* Roxb., and some species of eucalypts (Das, 2014; Locher and Currie, 2010; McGookin and Heilbron, 1926; Nair, 1995, 2000; Subrahmanyam, 1981; Subrahmanyam and Shah, 1988). Although chemically classified as gum, i.e. a mixture of polysaccharides, kino differs from it by having polyphenols that make kino astringent, contrary to gum (Badkhane *et al*., 2010; Hillis, 1975). The polyphenols are mostly tannins, essentially of flavonoid nature, typically making up 60 to more than 90 % (Bolza, 1978; Locher and Currie, 2010; Maiden, 1890a, b, 1892; Martius *et al*., 2012; Watt and Breyer-Brandwijk, 1962). Among these, kinotannic acid, which can make up to 75% (Setia, 1976), kinoin, kino red, and catechol are the main components of kino found in many species, e.g., *Acacia nilotica* (L.) Delile (Ali *et al*., 2012), *Eucalyptus camaldulensis* Dehnh. (Watt and Breyer-Brandwijk, 1962), and *Pterocarpus marsupium* Roxb. (Badkhane *et al*., 2010). When exposed to air the sticky straw-coloured kino solidifies and becomes a red to brown or black brittle semi-transparent solid, depending on the species and age of the plant. Enzymatic activity and the simultaneous evaporation of its volatile components are thought of causing kino solidification (Penfold, 1961).

Some gummous components of kino are synthesized within Golgi bodies and transported by the Golgi-mediated smooth vesicles depositing their contents between the plasmalemma and the cell wall. Consequently, the cytoplasm contracts as more secreted substances accumulate and withdraws from the cell wall (Fahn, 1988a, b; Gedalovich and Fahn, 1985a; Nair *et al*., 1983; Subrahmanyam and Shah, 1988). Other gummous components, e.g. polysaccharides, are released into the vein lumen by sloughing off the outer layers of the epithelium cell wall (Setia, 1984b; Venkaiah, l986), similarly to what happens in gum ducts (Fahn and Evert, 1974; Nair, 1995; Venkaiah, 1992). The polyphenols are synthesized by the endoplasmic reticulum and then transported and released into the vacuole through the vesicle transfer system (Catesson and Moreau, 1985; Shahidi and Yeo, 2016). Vesicles released from it fuse with the plasmalemma releasing their contents between the plasmalemma and the cell wall (Nair *et al*., 1983). Only after the degradation of the cell wall, the autolysed protoplast and all secreted substances that make kino are released into the vein lumen (Nair, 1995; Nair *et al*., 1983).

## 1.5. The physiological role of kino

Together with axial parenchyma, lack of fibres, and few conducting elements, kino veins are considered a resistance strategy from plants by acting as a barrier zone protecting the cambium and preventing access of boring insects, bacteria, and fungi (Eyles and Mohammed, 2002; Hillis, 1987; Rajput *et al*., 2005; Tippett and Shigo, 1981, Wilkes, 1986). Infected plants defend themselves and resist the spread of pathogens by compartmentalization, i.e. forming boundaries that isolate the injured or necrotic tissues from the living cambium. It starts with the accumulation of chemicals, e.g. polyphenols, gum, within tissues as the plant and pathogen interact and progresses to the formation of both chemical and anatomical boundaries, as the cambium responds by forming a barrier zone between the infected and the new tissues (Babu and Shah, 1987; Shigo, 1984). Therefore, barrier zones are important in restricting the transport of either fungal toxins or phytotoxic metabolites produced by host necrotic reactions towards the cambium. Kino astringency is thought of related to antimicrobial activity, as some of its phenolic components have shown antibacterial, antiviral, and fungicidal properties (Locher and Currie, 2010; Martius *et al*., 2012).

## 2. Kino flow model

Let one adapt the model approach of Cabrita (2018) to a fully developed kino vein of cylindrical geometry with lumen radius *R*, surrounded by an active and stimulated epithelium that autolyses (Fig. 3). The vein is open at the end of its length *L* where kino, of constant viscosity $\mu$ and density $\rho$ (i.e. an incompressible fluid), flows out of it through a wound, and solidifies when exposed to the outside air at normal pressure and temperature. For the sake of simplicity and mostly due to lacking of experimental data, one neglects the contribution of the balance between the accumulation of gummous and polyphenolic compounds, autolysis, and meristematic activity of the epithelium to the enlargement of the vein lumen, as described before, by considering the kino vein of constant radius *R* (Fig. 3). A higher rate of cell autolysis than cell division contributes to the enlargement of the kino vein lumen, i.e. an increase of its radius *R*. However, if both processes balance themselves out, *R* will remain constant (Fig. 3). Considering kino veins dimensions (Dowden and Foster, 1973; Eyles and Mohammed, 2003; Fahn, 1990; Greenwood and Morey, 1979; Skene, 1965; Tippett, 1986), kino flow velocity (Bhatt and Mohan Ram, 1990; Hillis and Hasegawa, 1963; Tripathi *et al*., 2015; Vasishth and Guleria, 2017), and viscosity (Hillis, 1962, 1987), all the necessary conditions are thus observed so that one can classify kino flow as Stokes flow. That is, kino flow is characterized by a very small Reynolds number[2], *Re*, between $10^{-2}$ and $10^{-1}$, similarly to what is observed on resin flow in conifers (Cabrita, 2018). Therefore, one can apply the unsteady Stokes equation (Cabrita, 2018) to describe the flow of kino within a vein with velocity *u*:

---

[2] $Re = \frac{uL}{\nu}$

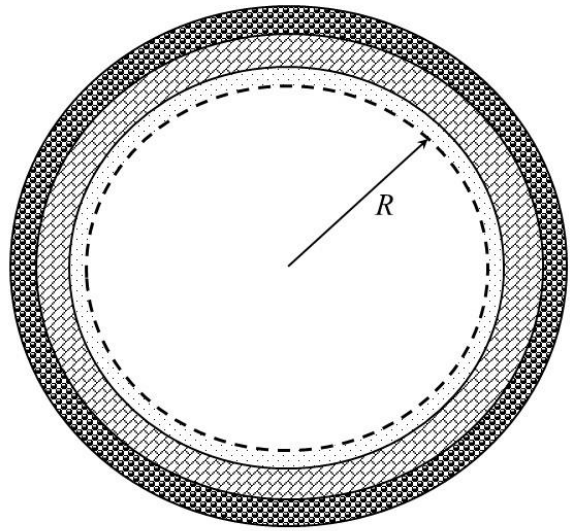

**Figure 3**

$$\frac{\partial \vec{u}}{\partial t} = -\frac{1}{\rho}\vec{\nabla}p + \nu\nabla^2 \vec{u} \qquad (1)$$

where $\nu = \mu/\rho$ is the kino kinematic viscosity and $p$ stands for the kino vein dynamic pressure, from now on simply called pressure, which arises wholly from the dynamic effects of the motion of kino (Batchelor, 2000; Kundu and Cohen, 2008). The dynamic pressure, includes the gravitational effects; i.e. $p = p' - p_0 - \rho g h$, where: $p'$ is the total pressure inside the kino vein lumen; $g$ the local acceleration of gravity; and $h$ is the vertical coordinate above a standard plane of reference where the hydrostatic pressure is $p_0$. The term $p_0 + \rho g h$ is the hydrostatic pressure that would exist in the kino vein if there is no flow. During kino flow, the hydrostatic pressure gives rise to a pressure gradient in balance with the force of gravity.

Similarly to what is observed on resin flow (Cabrita, 2018), the rate of kino solidification at the wound, $z = L$, (m$^3$.s$^{-1}$) is much smaller than the rate of kino outflow at that point. Therefore, the closing of the wound and the subsequent local decrease of the kino vein radius $R$ happen at a much slower pace and do not change the flow dynamics, i.e. kino flow remains laminar. For this reason, considering the kino veins dimensions, specifically that $L/R \gg 1$, through which laminar flow occurs, and the fact that the solidification of kino happens mostly outside on the bark, the end effects on the flow caused by changes in the vein lumen resistance at the wound, arisen mostly by the local decrease in $R$, can be neglected. As an incompressible fluid, applying the continuity equation to kino flow implies that:

$$\nabla \cdot \vec{u} = 0 \qquad (2)$$

## 2.1. Boundary and initial conditions

a) **Kino vein lumen boundary, $r = R$**: let one assume that the influx of kino, $u_r(R,z,t)$, is time-dependent only and linear with the rate of autolysis of epithelial cells of the peripheral cambium, $\frac{dN_e}{dt}$ (Fig. 3), through which kino is released into the vein lumen after the accumulation of polyphenolic compounds in the vacuole and gummous components between the plasmalemma and the cell wall. Therefore, considering the nature of holocrine secretion, described before, the loading of kino into the vein lumen occurs similarly throughout its entire length at a rate that is constant with the axial distance that defines the vein axis, $z$. That is:

$$u_r(R,z,t) \propto \frac{dN_e}{dt} \qquad (3)$$

where $N_e$ is the number of autolysing epithelial cells. Hence, in these conditions, the radial velocity component $u_r$ is a function of the radial distance $r$ and time $t$ solely, and therefore assumed of the form:

$$u_r(r,z,t) \equiv \xi(r)\theta(t) \qquad (4)$$

where $\xi$ and $\theta$ are undetermined functions. At the kino vein lumen boundary, which is considered fixed, there is the no-slip condition:

$$u_z(R,z,t) = 0 \qquad (5)$$

Which means that in this region the flow velocity, $u$, is in the radial direction only, reflecting the holocrine loading of kino that occurs similarly throughout the entire length of the kino vein (3).

**b) Centre of the kino vein, $r = 0$**: the cylindrical symmetry of the kino vein (Fig. 3) implies that:

$$u_r(0, z, t) = 0 \tag{6}$$

$$\frac{\partial u_z}{\partial r}(0, z, t) = 0 \tag{7}$$

and

$$\frac{\partial p}{\partial r}(0, z, t) = 0 \tag{8}$$

At the centre of the vein lumen, pressure, $p$, and flow velocity, $u$, vary only with time and the axial distance, $z$.

**c) Origin, $z = 0$**: at $t = 0$ when the vein is ruptured one has that:

$$\bar{u}_z(0,0) = \frac{2}{R^2} \int_0^R r u_z(r,0) \, dr = U_0 \tag{9}$$

and

$$\bar{p}(0,0) = \frac{2}{R^2} \int_0^R rp(r,0)dr = p_i \tag{10}$$

where $U_0$ is the average initial axial velocity, $p_i$ is the average initial pressure, and $\bar{u}_z$ and $\bar{p}$ are the average axial velocity and pressure in the vein respectively.

d) **Wound region, $z = L$**: at $t = 0$, when the vein is ruptured:

$$\bar{u}_z(L,0) = U \tag{11}$$

and

$$\bar{p}(L,0) = p_a \tag{12}$$

There is outflow of kino that is subsequently solidified once in contact with the air outside. That is, the flow of the liquid fraction of kino being exuded through the wound (m³.s⁻¹) is given by:

$$\frac{dV}{dt} = \iint u_z(r,L,t) r\,dr\,d\varphi - \delta \tag{13}$$

So that at $t = 0$, there is no volume of kino exuded yet to be collected. That is:

$$V(0) = 0 \tag{14}$$

Being *V* the volume of the liquid fraction of the kino exuded at the wound, with an average initial outflow velocity *U*, where the average pressure, $\overline{p}(L,0)$, equals the local atmospheric pressure, $p_a$. Part of the kino exuded solidifies at a rate $\delta$ (m$^3$.s$^{-1}$), which is assumed constant for modelling purposes.

All symbols used are defined in the text and in Table 2. Due to the cylindrical symmetry of the kino vein lumen to its axis, the dependent variables velocity, *u*, and pressure, *p*, are independent of the angular coordinate, the azimuthal angle $\phi$, in what is defined as axisymmetric flow (Batchelor, 2000). Hence, using definition (4) and the continuity Eq. (2) one concludes that the axial velocity $u_z$ is a linear function of the axial distance *z*, and the scalar governing equations in cylindrical coordinates derived from the unsteady Stokes equation describing flow within the kino vein (1) become:

$$\frac{\partial u_r}{\partial t} = -\frac{1}{\rho}\frac{\partial p}{\partial r} + \nu \frac{\partial}{\partial r}\left[\frac{1}{r}\frac{\partial (ru_r)}{\partial r}\right] \tag{15}$$

$$\frac{\partial u_z}{\partial t} = -\frac{1}{\rho}\frac{\partial p}{\partial z} + \frac{\nu}{r}\frac{\partial}{\partial r}\left(r\frac{\partial u_z}{\partial r}\right) \tag{16}$$

## 2.2. Properties of kino flow

Solving governing Eqs. (15) and (16) considering the boundary conditions described before (3) to (12) (see "Appendix"), the axial and radial velocity components and the average axial velocity are given by respectively:

**Table 2** – List of symbols used

| Symbol | Parameter | SI units |
|---|---|---|
| $\delta$ | Rate of kino solidification | $m^3.s^{-1}$ |
| $L$ | Kino vein length | m |
| $\mu$ | Kino viscosity | Pa.s |
| $N_e$ | Number of autolysing epithelial cells | –– |
| $\nu$ | Kino kinematic viscosity | $m^2.s^{-1}$ |
| $p$ | Pressure within the kino vein | Pa |
| $\bar{p}$ | Average pressure within the kino vein | Pa |
| $p_a$ | Local atmospheric pressure | Pa |
| $p_i$ | Initial average pressure within the kino vein | Pa |
| $r$ | Radial coordinate | m |
| $R$ | Kino vein radius | m |
| $\rho$ | Kino density | $kg.m^{-3}$ |
| $\sigma$ | Kino vein specific resistance | $Pa.s.m^{-2}$ |
| $t$ | Time | s |
| $t_c$ | Time at which the wound closes and kino stops flowing due to the accumulation of solidified kino at the wound | s |
| $u$ | Velocity | $m.s^{-1}$ |
| $u_r$ | Radial velocity | $m.s^{-1}$ |
| $u_z$ | Axial velocity | $m.s^{-1}$ |
| $\bar{u}_z$ | Average axial velocity | $m.s^{-1}$ |
| $U$ | Initial average kino outflow velocity at the wound | $m.s^{-1}$ |
| $U_0$ | Initial average axial velocity | $m.s^{-1}$ |
| $V$ | Volume of liquid fraction of kino exuded at the wound | $m^3$ |
| $z$ | Axial coordinate | m |

$$u_z(r,z,t) = \frac{(U-U_0)\left(J_0\left(\frac{\sqrt{K}}{R}r\right) - J_0\left(\sqrt{K}\right)\right)}{L\left(\frac{2}{\sqrt{K}}J_1\left(\sqrt{K}\right) - J_0\left(\sqrt{K}\right)\right)}\left(z + \frac{U_0 L}{U-U_0}\right)e^{-\frac{K\nu}{R^2}t} \quad (17)$$

$$u_r(r,z,t) = \frac{(U-U_0)\left(\frac{r}{2}J_0\left(\sqrt{K}\right) - \frac{R}{\sqrt{K}}J_1\left(\frac{\sqrt{K}}{R}r\right)\right)}{L\left(\frac{2}{\sqrt{K}}J_1\left(\sqrt{K}\right) - J_0\left(\sqrt{K}\right)\right)}e^{-\frac{K\nu}{R^2}t} \quad (18)$$

and

$$\bar{u}_z(z,t) = \left(\frac{U-U_0}{L}z + U_0\right)e^{-\frac{K\nu}{R^2}t} \quad (19)$$

where $J_0$ and $J_1$ are Bessel functions of zero and first orders, respectively, and K is a non-dimensional undetermined parameter characterizing kino loading into the vein, regardless of its dimensions, which due the nature of holocrine loading of kino it might depend on the species. From result (18), one has that the holocrine loading of kino at the vein lumen boundary, $r = R$, (4) (Fig. 3) is given by:

$$u_r(R,z,t) = -\frac{(U-U_0)R}{2L}e^{-\frac{K\nu}{R^2}t} \quad (20)$$

Solutions (17) and (18) show a typical time dependence of the unsteady Stokes flow regime, which, in this case, evolves by decaying exponentially with time to a steady-state situation when flow stops once the wound at the end of the kino vein, $z = L$, is closed by solidified kino (13). Hence, $\frac{K\nu}{R^2}$ is the decay constant, of SI units s$^{-1}$, that equals the fractional change of the loading of kino with time:

$$\frac{K\nu}{R^2} = -\frac{\frac{du_r}{dt}(R,z,t)}{u_r(R,z,t)} \tag{21}$$

Therefore, based on these results, the half-life, $t_{1/2}$, of kino flow, i.e. the time required for it to reduce to half of its initial value, is:

$$t_{1/2} = \frac{\ln(2)R^2}{K\nu} \tag{22}$$

Physiologically, one can interpret the decay constant $\frac{K\nu}{R^2}$ as describing the rate of holocrine secretion of kino into a fully developed vein resulting from cell lysigeny that occurs due to stimulus and happens throughout its length (Fig. 3). From result (20), one has that higher kino loading rates are kept through time, i.e. changing more slowly with time for smaller values of the decay constant $\frac{K\nu}{R^2}$, in veins with smaller surface-area-to-volume ratio[3]; that is, wider and longer veins. Therefore, wider veins present higher values of kino flow half-life (22). This results is expected, considering that holocrine loading will be higher in wider kino veins surrounded by a larger autolysing epithelium than in narrower kino veins surrounded by smaller epithelia; thus with a smaller number of autolysing cells. For a given set of vein dimensions, the smaller value of the decay constant reflecting higher holocrine loading rates is achieved by having smaller values of parameter K. In this case, higher holocrine loading rates may be expressed either by faster accumulation of polyphenolic and gummous components or cell autolysis or even both. In addition, a stronger meristematic activity of the epithelium affecting these processes will also contribute to a higher holocrine loading of kino. The synthesis of less viscous and denser kinos further contributes to smaller value of the decay constant (21), i.e. higher holocrine loading rates (20).

---

[3] From the surface-area-to-volume ratio of a cylinder of radius $R$ and length $L$, $SA:V = \frac{2(L+R)}{LR}$, one sees that it decreases for longer and wider cylindrical pathways.

From the continuity equation, i.e. mass conservation (2), or, alternatively, from Eqs. (19) and (20) one finds that:

$$\frac{\partial \overline{u}_z}{\partial z} = -\frac{2}{R} u_r(R,z,t) = \frac{(U-U_0)}{L} e^{-\frac{K\nu}{R^2}t} \qquad (23)$$

The change in the average flow with distance is linear with the holocrine loading of kino, i.e. flow increases towards the wound as more kino is loaded into the vein lumen, due to epithelial cells autolysis (3). However, the magnitude of the changes in the average flow with distance decreases with time reflecting the concomitant decrease in holocrine loading of kino (20).

Pressure in the kino vein is given by:

$$p(r,z,t) = p_i + \frac{(U-U_0)\mu K}{LR^2\left(\frac{2J_1(\sqrt{K})}{\sqrt{K}J_0(\sqrt{K})}-1\right)}\left[\frac{2r^2-R^2}{8}-\frac{(U-U_0)z^2+2U_0Lz}{2(U-U_0)}\right]e^{-\frac{K\nu}{R^2}t} \qquad (24)$$

and the average pressure within it is:

$$\bar{p}(z,t) = p_i - \frac{\mu K \left[(U-U_0)z^2 + 2U_0 Lz\right]}{2LR^2 \left(\frac{2J_1(\sqrt{K})}{\sqrt{K}J_0(\sqrt{K})} - 1\right)} e^{-\frac{K\nu}{R^2}t} \qquad (25)$$

From this result one has that the average pressure gradient is a linear function of the axial distance $z$:

$$\frac{\partial \bar{p}}{\partial z} = -\frac{\mu K \left[(U-U_0)z + U_0 L\right]}{LR^2 \left(\frac{2J_1(\sqrt{K})}{\sqrt{K}J_0(\sqrt{K})} - 1\right)} e^{-\frac{K\nu}{R^2}t} \qquad (26)$$

as it is expected when the average pressure presents a parabolic profile with the axial distance $z$ (25). Comparing it with the average velocity (19) one concludes that:

$$\frac{\partial \bar{p}}{\partial z} = -\frac{\mu K}{R^2 \left(\frac{2J_1(\sqrt{K})}{\sqrt{K}J_0(\sqrt{K})} - 1\right)} \bar{u}_z \qquad (27)$$

The average pressure gradient is linear with the average axial velocity as it is expected for a laminar flow regime (Cabrita, 2018; Kundu and Cohen, 2008) in which the kino vein specific resistance $\sigma$ (Pa.s.m$^{-2}$) is given by:

$$\sigma = \frac{\mu K}{R^2 \left( \dfrac{2J_1\left(\sqrt{K}\right)}{\sqrt{K} J_0\left(\sqrt{K}\right)} - 1 \right)} \tag{28}$$

Using this result, the average pressure in the kino vein (25) can be written as:

$$\overline{p}(z,t) = p_i - \frac{\sigma}{2L}\left[(U - U_0)z^2 + 2U_0 Lz\right] e^{-\frac{K\nu}{R^2}t} \tag{29}$$

and the change in the average pressure with time is given by:

$$\frac{\partial \overline{p}}{\partial t} = \frac{\nu K \sigma}{2LR^2}\left[(U - U_0)z^2 + 2U_0 Lz\right] e^{-\frac{K\nu}{R^2}t} \tag{30}$$

From result (29) one finds that the decline in the kino vein pressure with distance in the direction of flow (27) will be smaller for veins presenting smaller values of the specific resistance, $\sigma$ (28). While pressure in wider kino veins presenting smaller specific resistance and higher holocrine loading rates, i.e. smaller values of the decay constant (21) favoured by smaller values of parameter K, will increase more slowly with time (30) allowing flow to continue for longer, as reflected in their higher values of kino flow half-life (22). This result agrees with the fact that wider kino veins delivering higher amounts of kino at the wound will take longer to close through the solidification of kino, which means that pressure in them will increase at a slower rate as the wound closes. Additionally, from result (30), one has that a slower change of pressure with time will be further enhanced by producing less viscous and denser kinos; that is, presenting smaller values of the kinematic viscosity, $\nu$.

Considering result (28) and the range of values observed on the dimensions of fully developed and functional kino veins and pockets (Dowden and Foster, 1973; Eyles and Mohammed, 2003; Fahn, 1990; Greenwood and Morey, 1979; Loewus and Runeckles, 1977; Skene, 1965; Tippett, 1986), one finds that the minimum value of parameter K possible, i.e. for which the kino veins offer resistance to flow, is $2.16 \times 10^{-8}$ approximately. Parameter K characterizes the holocrine loading of kino into kino veins and does not depend on their dimensions rather on the physiology (3); i.e. on the rates of accumulation of gummous components between the plasmalemma and the cell wall and phenolic compounds in the vacuole, which are all released into the vein lumen upon subsequent cell autolysis of the inward lining layers of the peripheral cambium (epithelium) (Fig. 3). The contribution and importance of these processes may vary among species, and depend on the age of the plant and stimuli given to vein formation. This might explain the likely variability of parameter K between species, and possibly individuals. Comparing the specific resistance of a kino vein of radius $R$ (28) with that of a tube of similar dimensions under Poiseuille flow regime, $\sigma_{\text{Poiseuille}}$, one finds that:

$$\frac{\sigma}{\sigma_{\text{Poiseuille}}} = \frac{K}{8\left(\dfrac{2J_1\left(\sqrt{K}\right)}{\sqrt{K}J_0\left(\sqrt{K}\right)} - 1\right)} \tag{31}$$

As Fig. 4A shows, the specific resistance of a kino vein, $\sigma$, is higher than that of a tube of similar dimensions under Poiseuille flow regime for smaller values of parameter K (31), i.e. for veins with higher holocrine loading rates of kino. Therefore, veins with smaller values of the decay constant

(21) have a higher specific resistance (28) (Fig. 4B). The kino vein specific resistance, $\sigma$, tends to $\sigma_{\text{Poiseuille}}$, its smallest value, i.e. $\frac{\sigma}{\sigma_{\text{Poiseuille}}} \approx 1$, as parameter K increases. That is, kino veins with higher values of the decay constant (21), hence with smaller holocrine loading rates of kino (20), present smaller resistance to kino flow (Fig. 4A). Smaller values of the decay constant represent higher holocrine loading rates that can be caused by higher rates of synthesis and accumulation of polyphenols and gummous components or cell autolysis or both, being all related to the epithelium meristematic activity. In this scenario, the kino veins specific resistance increases as the values of parameter K decrease (Fig. 4A). This means that for delivering flow at a given velocity higher pressure gradients are needed for veins presenting higher specific resistance (28) for smaller values of parameter K, as result (27) indicates. Hence, to build up such pressure gradients, higher kino loading rates are needed (20). Computing the kino vein specific resistance, $\sigma$ (28), for different values of parameter K for the range of values of the lumen radius, $R$, observed for kino veins and pockets, one sees that it decreases sharply and hyperbolically with $R^2$ not changing much between different vein dimensions for $K > 5 \times 10^{-7}$ approximately (Fig. 4B). Increasing the value of parameter K up to two orders of magnitude does not change this scenario. That is, the increase in the specific resistance, $\sigma$, due to the decrease in the values of parameter K (Fig. 4A), seems to be balanced out by plants by investing in having wider kino veins (Fig. 4B), which ultimately leads to maintaining smaller pressure gradients needed to drive flow.

Integrating Eq. (13), considering boundary conditions (11), (14), and result (17), the volume of the liquid fraction of kino being exuded at the wound is given by:

$$V(t) = \frac{U\pi R^4}{K\nu}\left(1 - e^{-\frac{K\nu}{R^2}t}\right) - \delta t \qquad (32)$$

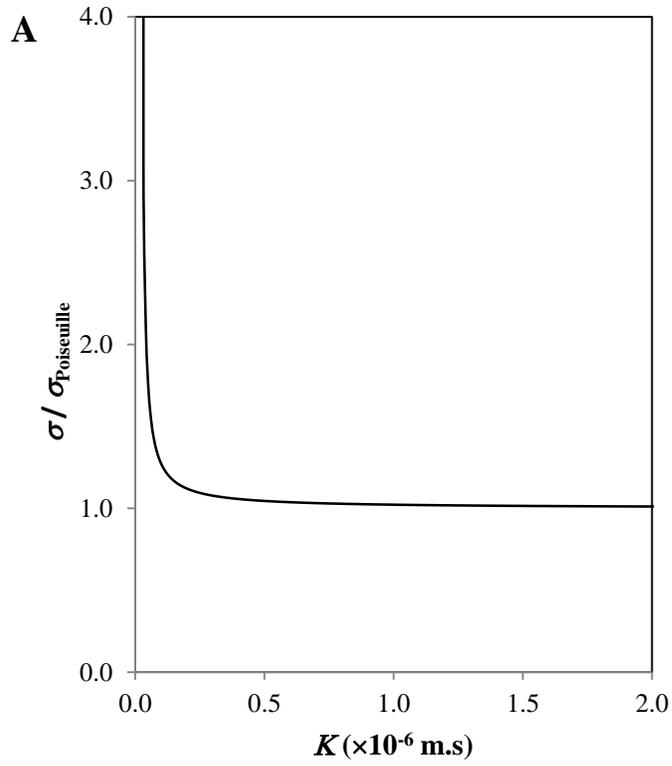
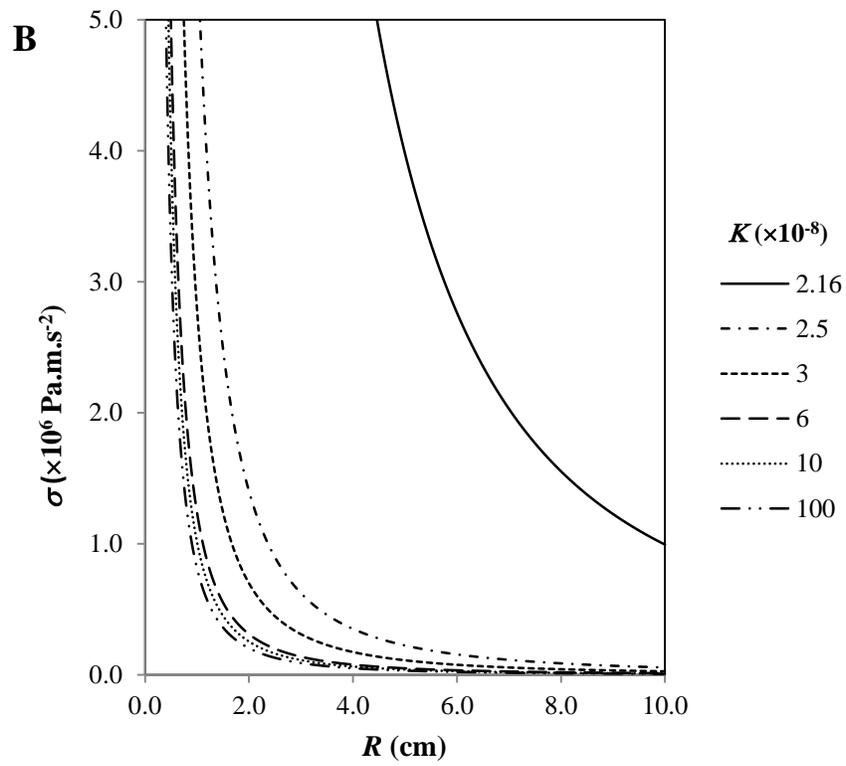

**Figure 4**

As suggested by the differences in pressure and kino loading between wider and narrower veins described before, wider kino veins with smaller values of parameter K, i.e. with smaller specific resistance (28) and presenting higher kino loading rates (20), thus having smaller values of the decay constant (21), are the ones delivering higher volume of kino at the wound, $z = L$. This is further enhanced for less viscous and denser kinos; i.e. with smaller values of the kinematic viscosity, $\nu$. Kino is subsequently solidified at a constant rate $\delta$. Eventually, the wound closes and the volume of exudate no longer changes with time beyond:

$$t_c = -\frac{R^2 \ln\left(\dfrac{\delta}{U\pi R^2}\right)}{\mathrm{K}\nu} \quad (33)$$

As expected, wider veins with smaller values of parameter K, thus delivering larger volumes of kino (32), are also the ones taking longer to close, i.e. presenting higher values of $t_c$ (33), in agreement with their higher kino flow half-life (22).

Unfortunately, experimental data on kino flow are rare compared to other plant exudates such as latex, oleoresins, and some gums economically important. This fact is most certainly related to the limited economic importance of kino as a niche product with limited applications (Locher and Currie, 2010; Martius *et al.*, 2012). However, there are some measurements, sadly scattered in time, from which one can obtain valuable information and get a clearer picture of the physiology behind kino flow in angiosperms. Kino is referred to having thick consistency similar to that of honey (Hillis, 1962, 1987), with a specific gravity between 1.1 and 1.4 (Maiden, 1890a, b, 1892, 1901). Therefore, considering the typical high content of tannins in kino (Bolza, 1978), its viscosity $\mu = 10$ Pa.s, i.e. 10,000 times as viscous as pure water, and density $\rho = 1,400$ kg.m$^{-3}$ at normal ambient temperature, one finds kino kinematic viscosity $\nu = 7.14 \times 10^{-3}$ m$^2$.s$^{-1}$. In addition, from the several observations made on kino and gum flow in some species (Table 3), one has that the average kino flow is not faster than the resin flow observed on pines (Cabrita, 2018; Hodges *et al.*, 1977, 1981; Schopmeyer *et al.*, 1954), i.e. in the order of $10^{-5}$ to $10^{-4}$ m.s$^{-1}$. The model was then tested for the case of a 0.5 cm radius kino vein and a 3 cm radius kino pocket, each of 1 m length, considering: K = $1 \times 10^{-7}$, $p_a$ constant with time and equal to 1 atm (101,325 Pa); $U = 1 \times 10^{-4}$ m.s$^{-1}$; and $U_0 = 1 \times 10^{-6}$ m.s$^{-1}$. For this value of parameter K considered, the specific resistance (28) of the 0.5 cm radius kino vein and the 3 cm radius kino pocket is about 1.3 times as much as that of ducts with similar dimensions under Poiseuille regime (Fig. 4A), that is $4.07 \times 10^6$ and $1.13 \times 10^5$ Pa.s.m$^{-2}$, respectively. Taking the values of the kino kinematic viscosity, $\nu$, and parameter K considered, the 0.5 cm radius kino vein has a decay constant (21) of $2.86 \times 10^{-5}$ s$^{-1}$, while this parameter for the 3 cm radius kino pocket is $7.94 \times 10^{-7}$ s$^{-1}$; that is about 1/36 the value of the kino vein decay constant. This gives a kino half-life (22) of 6.7 h and 10.1 d for the 0.5 cm radius kino vein and the 3 cm radius kino pocket, respectively.

Table 3 – Average mass flux obtained from experimental data on species that present holocrine secretion in their internal secretory systems.

| Species | Exudate | Mass flux ($\times 10^{-5}$ kg.s$^{-1}$.m$^{-2}$) | Source of data |
|---|---|---|---|
| *Anogeissus latifolia* (Roxb. ex DC.) Wall. ex Guill. & Perr. | Gum | 0.164 – 193 | Bhatt (1987); Ghritlahare (2017); Kuruwanshi *et al*. (2018) |
| *Buchanania cochinchinensis* (Lour.) M.R.Almeida | Gum | 5.13 – 22.6 | Ghritlahare (2017) |
| *Firmiana simplex* (L.) W.Wight | Gum | 0.437 – 198 | Nair (2003); Nair *et al*. (1995); Kuruwanshi *et al*. (2017) |
| *Lannea coromandelica* (Houtt.) Merr. | Gum | 1.21 – 30.5 | Vasishth (2017) |
| *Soymida febrifuga* (Roxb.) A.Juss. | Gum | 8.53 – 89.4 | Ghritlahare (2017) |
| *Terminalia tomentosa* Wight & Arn. | Gum | 3.35 – 6.84 | Ghritlahare (2017) |
| *Acacia nilotica* (L.) Delile | Kino | 0.114 – 29.6 | Das (2014); Raj (2015); Kuruwanshi (2017) |
| *Acacia senegal* (L.) Willd. | Kino | 0.780 – 31.0 | Bhatt and Mohan Ram (1990); Vasishth and Guleria (2017) |
| *Azadirachta indica* A.Juss. | Kino | 0.068 – 0.580 | Das (2014) |
| *Butea monosperma* (Lam.) Taub. | Kino | 0.031 – 0.151 | Tripathi *et al*. (2015) |
| *Moringa oleifera* Lam. | Kino | 0.068 – 1.34 | Das (2014) |
| *Eucalyptus sieberi* L. A.S.Johnson | Kino | 0.347 – 9.22 | Hillis and Hasegawa (1963) |

Having a time-dependent holocrine loading of kino that does not change with the axial distance (20) results in having a parabolic decrease of pressure (25) within kino veins in the direction of flow (Fig. 5A), which means that the pressure gradient (26) increases linearly in magnitude with distance *z* towards the wound (Fig. 5B). Consequently, the average flow velocity (19) increases linearly with the axial distance (Fig. 6A). Once in contact with the air outside, kino solidifies eventually sealing the wound completely. As more kino solidifies, partially blocking the exit wound, pressure at the wound and consequently throughout the vein lumen builds up, so that 48 h after rupture pressure within the 0.5 cm radius kino vein is almost uniform throughout (Fig. 5A). In this case, the average pressure gradient is practically non-existent (Fig. 5B), and flow (Fig. 6A) as well as loading of kino (Fig. 6B) have almost ceased completely. As times passes and the loading of kino decreases in magnitude (Fig. 6B), so does the average flow velocity (Fig. 6A) and the pressure gradient becomes less prominent, i.e. less negative (Fig. 5B). Naturally and in order to maintain a Stokes flow regime, pressure within kino veins evolves by decaying exponentially with time towards equilibrium (25), after which flow no longer occurs (Fig. 5). The closing of the wound by kino solidification just makes this scenario happening sooner (33). Essentially, the behaviour of kino flow is similar to that of resin flow observed on conifers (Cabrita, 2018). The different spatial and temporal profiles of kino flow dependent variables, i.e. pressure, flow velocity, and radial flow, arise only from the very different nature of the holocrine secretion of kino (3), compared to the granulocrine pressure-driven loading of resin (Cabrita, 2018). Similarly to resin flow, due to the holocrine loading of kino throughout the vein, the axial flow velocity has a varying quasi parabolic radial profile that sharpens towards the wound with the axial distance $z$, being maximum at the vein centre but decaying in magnitude with time (Fig. 7). This contrasts with the constant parabolic profile of the flow velocity typical of the Poiseuille flow regime, thus invalidating the applicability of Poiseuille equation to kino flow in the same way as observed on resin flow (Cabrita, 2018; Schopmeyer *et al.*, 1954). Qualitatively, the scenario here described for a 0.5 cm radius kino vein is not different from that of a 3 cm radius kino pocket. However, given its higher surface-area-to-volume ratio, both the kino pocket specific resistance (28) and decay constant $\frac{K\nu}{R^2}$ are smaller than that observed on the narrower kino vein. This means that the disturbance caused by rupturing the kino pocket at its end will remain longer through time, i.e. the pressure (Fig. 8A), pressure gradient (Fig. 8B), kino loading rate (Fig. 6B), and flow velocity (Fig. 9) will change more slowly with time than in the narrower kino vein (Figs. 5, 6A). Due to the kino pocket higher loading rate (20) (Fig. 6B), the change, i.e. the decrease, in the average axial velocity with time will be smaller (23) (Fig.

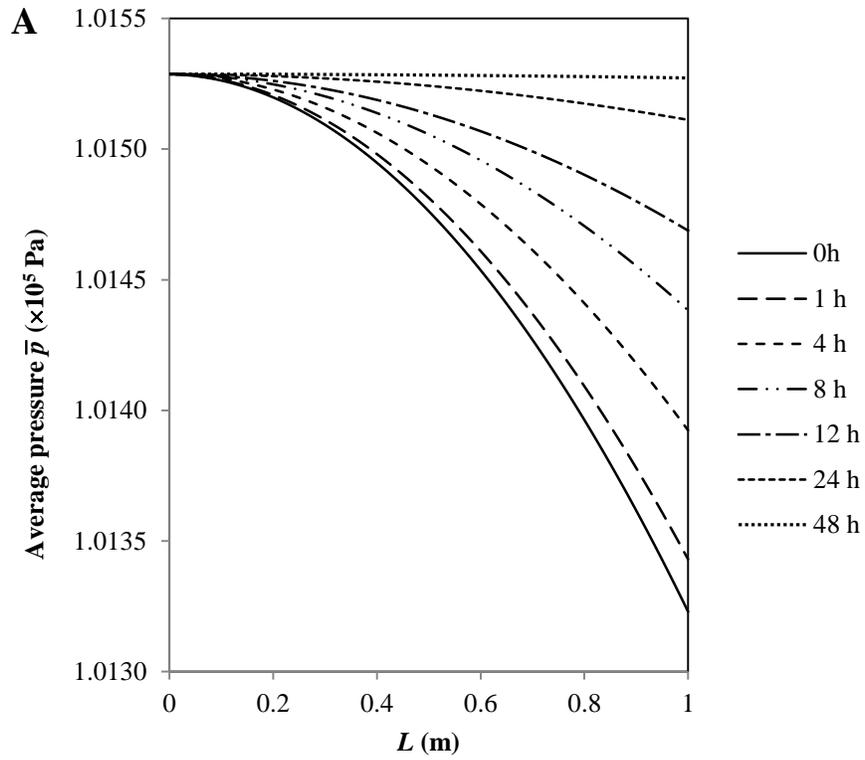

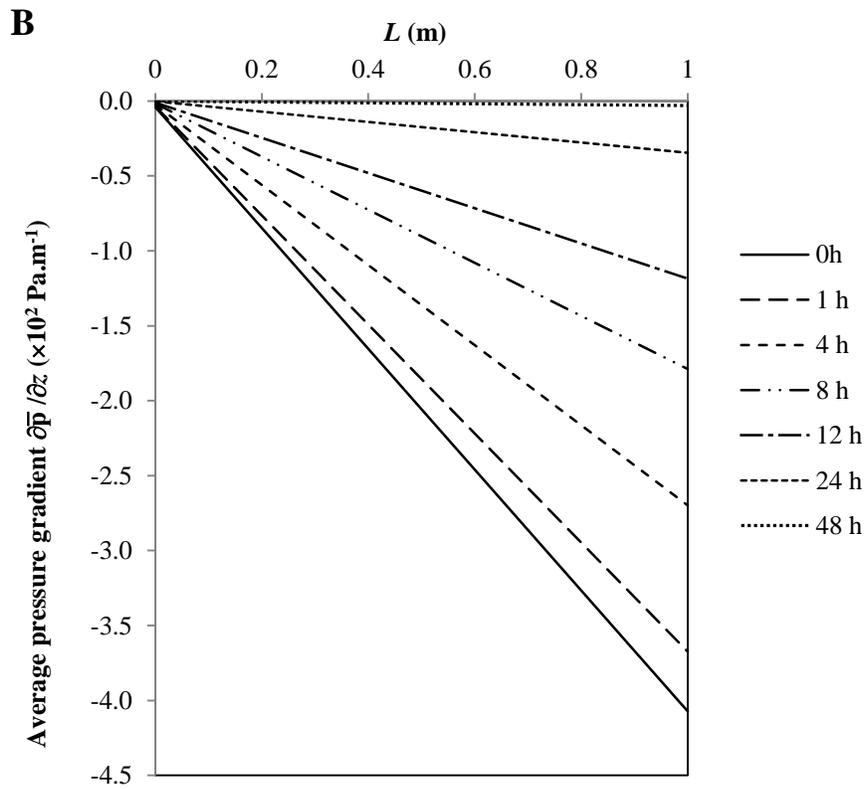

**Figure 5**

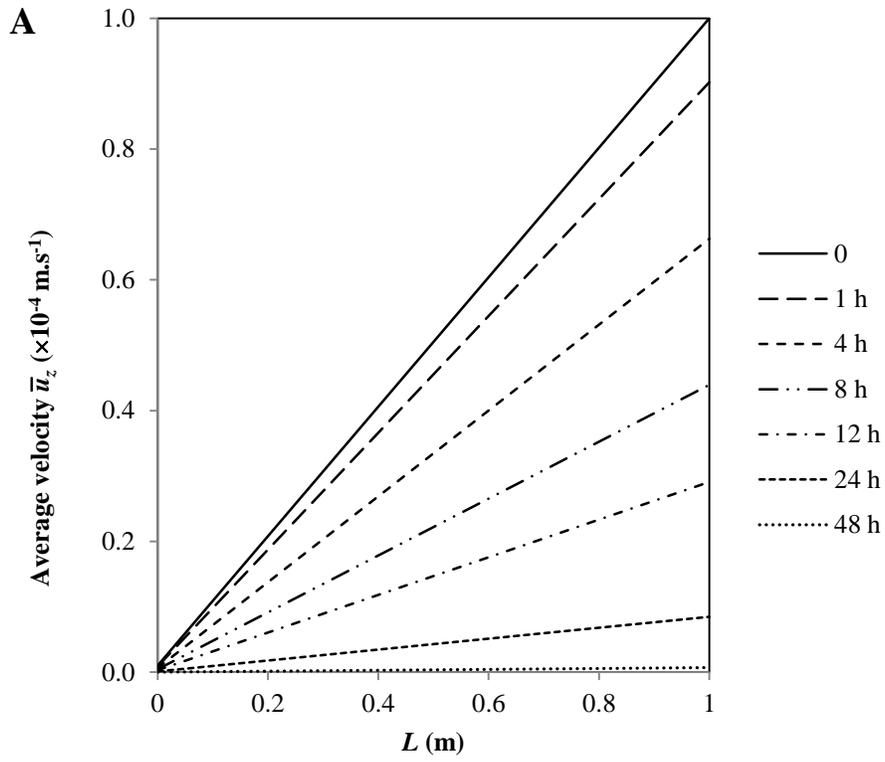

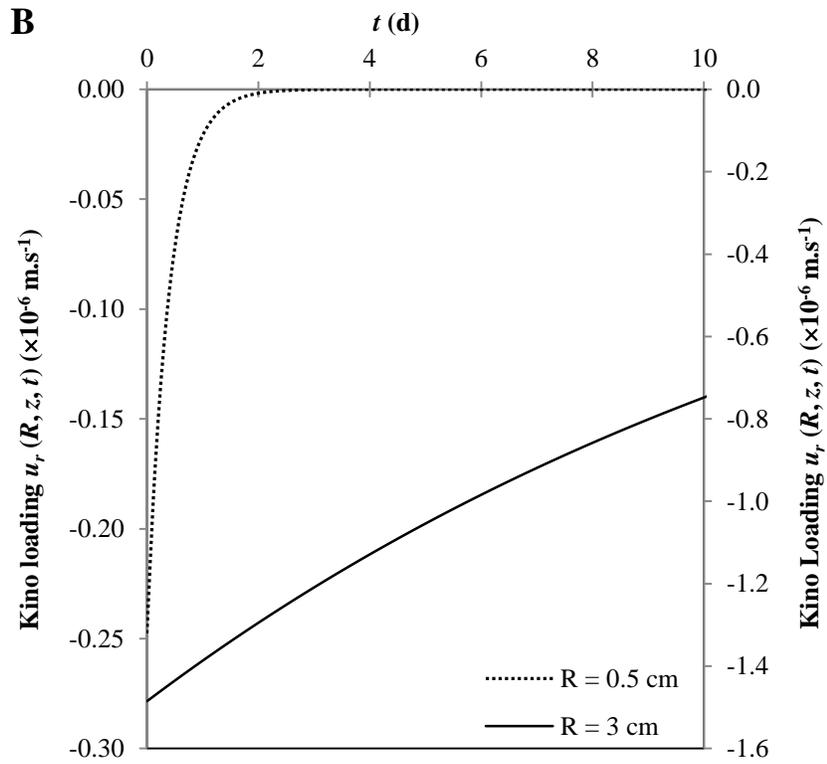

**Figure 6**

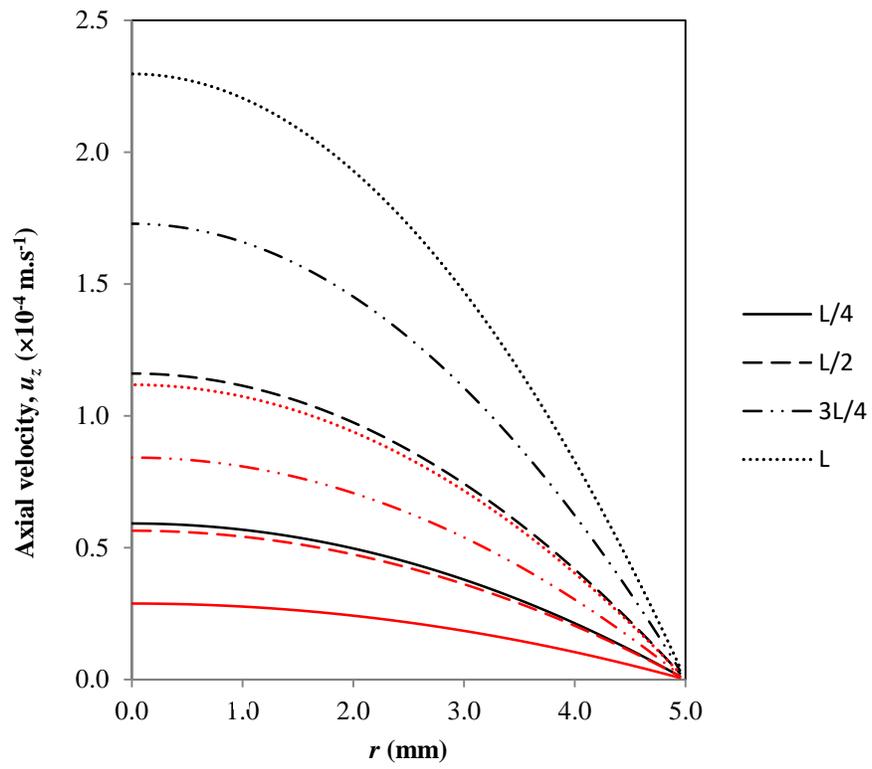

**Figure 7**

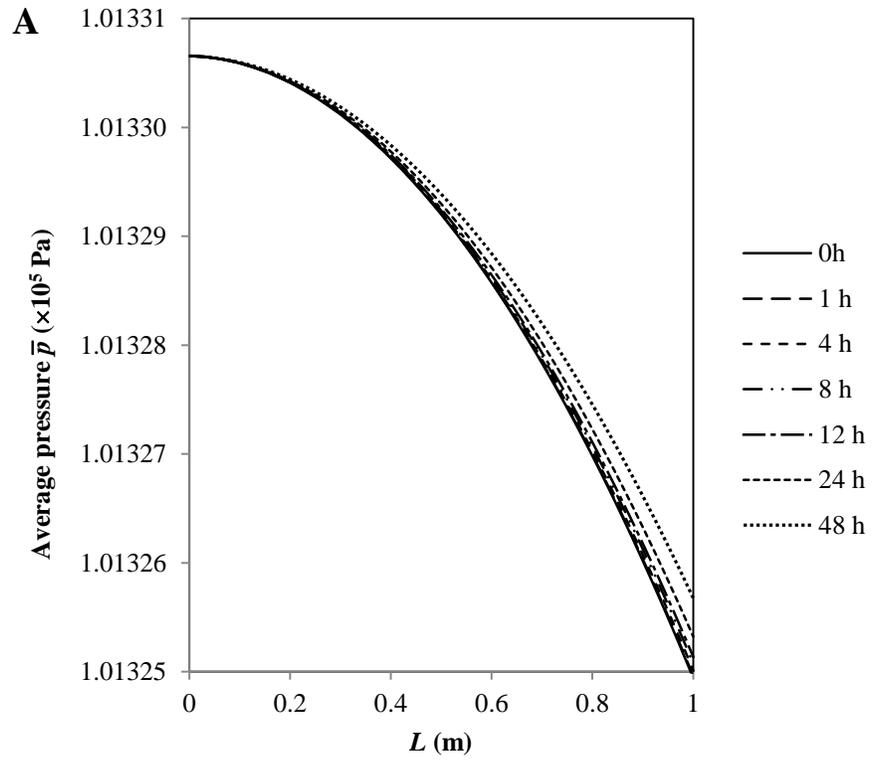

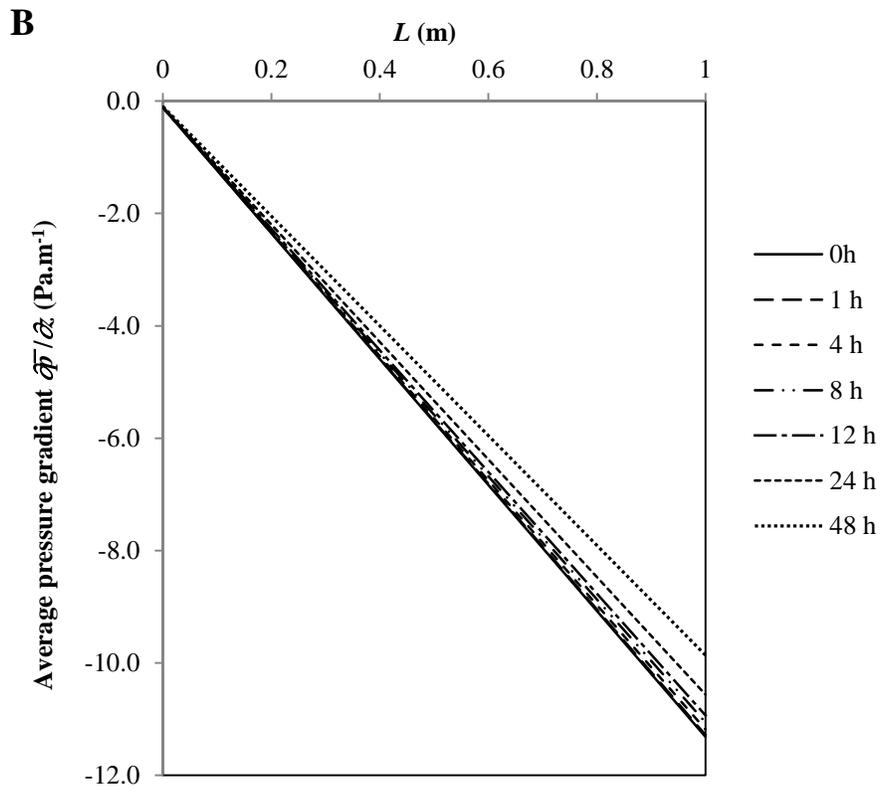

**Figure 8**

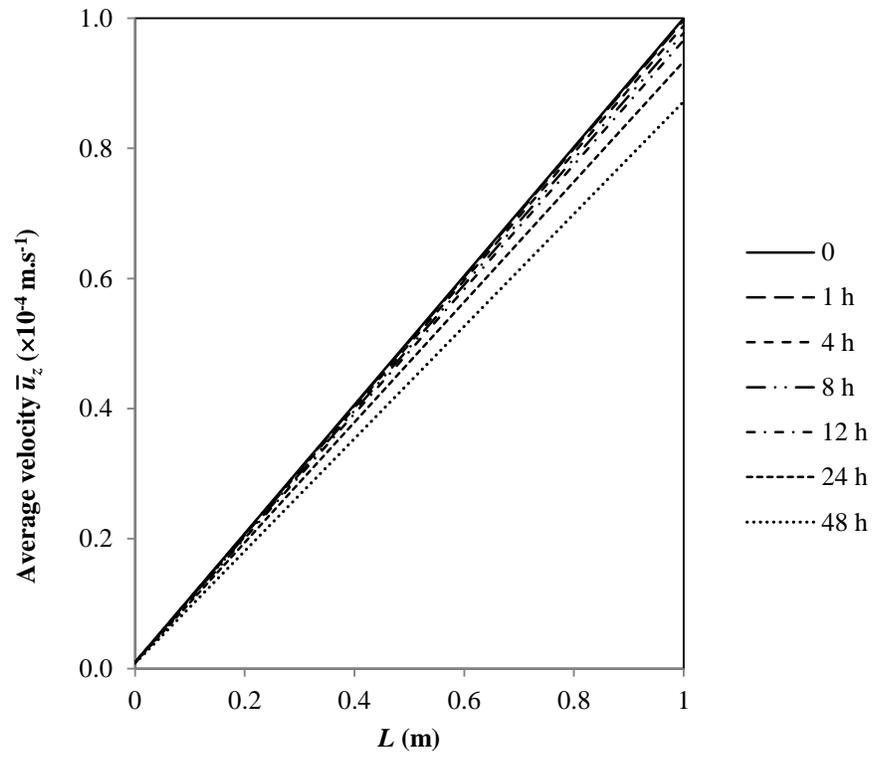

**Figure 9**

9). Additionally, due to its smaller specific resistance, pressure in the kino pocket will decrease less with the axial distance $z$ (29) (Fig. 8A). This means that compared to narrower veins, not only smaller pressure gradients are needed to drive flow throughout larger kino pockets, but also they are kept longer through time (Fig. 8B). Therefore, the holocrine secretion of kino into wider kino pockets allows them to deliver larger volumes for longer periods of time compared to narrower veins. As Fig. 6B shows, 48 h after rupture the loading of kino into a 3 cm radius kino pocket has decreased about 13% of its initial value, while in a narrower 0.5 cm radius vein it has practically ceased. About 7 kino flow half- lives have elapsed after 48 h in the narrower vein, whilst at that time only a 1/5 of the kino flow half-life has elapsed in the kino pocket. Consequently, at that time, while the kino pocket is still able to deliver flow at a velocity that is a bit less than 90% of what it was initially (Fig. 9), the average flow velocity in the narrower vein has become essentially non-existent (Fig. 6A), i.e. basically unmeasurable. Similarly to narrower veins (Fig. 7), kino flow within kino pockets also presents a varying quasi parabolic radial profile that sharpens towards the wound with the axial distance, decaying in magnitude with time, and reaching its maximum at the centre, $r = 0$. Flow stops (33) after $t_c = 67$ h in the narrower vein and $t_c = 100$ d in the kino pocket approximately, giving a kino solidification rate, $\delta$, in the order of $10^{-11}$ $m^3.s^{-1}$.

## 3. Discussion

Despite sharing many similarities in terms of their role in plant defence strategies, the internal secretory systems that rely on holocrine secretion, i.e. involving the lysis of the surrounding epithelium, here exemplified by kino veins, differ from those presenting granulocrine or eccrine secretion, e.g. resin ducts. The differences observed are not only in development, distribution, and structure, but also physiological, and consequently on the flow through these systems. Therefore, these differences might affect the performance of internal secretory systems used by plants for defence. However, regardless of being considered of limited occurrence in plants (Nair *et al*., 1983, 1995), holocrine secretion seems to present some advantages regarding the loading and flow of exudates inside ducts and cavities when compared to pressure-driven granulocrine secretion (Cabrita, 2018).

## 3.1. Kino loading is not pressure-driven and its flow requires less metabolic energy

Typically, kino veins form a tangential anastomosing network in the stem and branches encircling the plant, while resin ducts, despite being present in all plant organs, distribute vertically and radially making, in some cases, a three-dimensional network extending throughout the plant body. The bigger dimensions of kino veins (e.g. when they become kino pockets) contribute greatly to their specific resistance being four to five orders of magnitude smaller when compared to the specific resistance of the largest resin ducts found in Pinaceae. Furthermore, physiologically, while the resin duct specific resistance also depends on the duct wall permeability to resin, which in turn depends on the wall composition and thickness (Cabrita, 2018), the kino vein specific resistance depends on the holocrine loading of kino through parameter K (28), which does not involve any membrane transport. Hence, there are no metabolic energy costs associated with membrane carriers or transporters. This means that the kino vein capacity to deliver a given volume of kino (32) is mostly determined by the rate of epithelium autolysis, $\frac{dN_e}{dt}$, (3) that relates to the synthesis and accumulation of polyphenolic compounds and gummous components. Additionally, both these processes depend on the meristematic capacity of the peripheral cambium in substituting autolysed epithelial cells by producing new ones that will accumulate and release kino subsequently. The balance and contribution of these processes may well depended on the species, age of the plant, and the stimuli given to vein formation, i.e. trauma. Therefore, the anatomy, structure and physiology of kino veins, and associated epithelium suggest that the holocrine loading of kino is not driven by pressure contrary to what is observed on the resin ducts of conifers (Cabrita, 2018). When there is lysis of the epithelium, the pressure-driven loading and filling of the vein lumen (Büsgen and Münch, 1929; Cabrita, 2018) seems mostly unlikely. The lysigenous kino vein lumen, often of irregular shape and wider in one direction, is differently surrounded and limited by a mixture of non-living and living cells at different stages of development, with some being partly disintegrated and releasing their contents into it (Babu and Shah, 1987; ; Rajput et al., 2005). This means that, in this scenario, any pressure difference between the surrounding and partly autolysing epithelium and the duct lumen is consequently more difficult to be regulated or maintained in order to drive flow from the epithelium into the duct lumen. Due to the disintegration of their membranes, the autolysing epithelial cells lining the duct are no longer capable of keeping any pressure difference through their cell membranes, i.e. of directly building up pressure in the duct lumen in the same way as it happens in resin ducts (Cabrita, 2018). This agrees with the observations of Skene (1965) on *Eucalyptus obliqua* L'Hér., in which the secretory activity of kino veins is dependent on the continued activity of the peripheral cambium; thus quite different from what is known about the synthesis and flow of resin in conifers (Bannan, 1936; Cabrita, 2018), as well as gum and gum resin in some angiosperms, e.g. in the normal ducts of *Ailanthus excelsa* Roxb. (Babu *et al*., 1987).

Apart from the anatomical and physiological differences between resin-producing species, with granulocrine pressure-driven loading, and gum-producing species with holocrine loading, here exemplified by kino veins, perhaps, the most striking feature distinguishing them is the values of pressure and pressure gradients observed on kino veins and resin ducts. Pressure inside resin ducts falls within the range of values observed on plant cells, i.e. between 0.3 and 1 MPa (Bourdeau and Schopmeyer, 1958; Cabrita, 2018; Rissanen *et al.*, 2015; Schopmeyer *et al.*, 1954). The model presented in this work suggests that pressure within kino veins, in the order of $1\times10^5$ Pa, is most likely smaller than that observed on resin ducts but still higher than the pressure observed on the apoplast (Nobel, 1999). Therefore, considering the much smaller kino veins specific resistance and that kino flow (Fig. 4A) may be equally as fast as resin flow (Cabrita, 2018; Hodges *et al.*, 1977, 1981; Schopmeyer *et al.*, 1954), the pressure gradients needed to drive kino flow, eventually through considerable distances, range up to four to five orders of magnitude smaller, i.e. up to $10^2$ Pa.m$^{-1}$ (Figs. 5B, 8B), than the pressure gradients observed on resin ducts (Cabrita, 2018). Also, the pressure gradients driving kino flow are up to three orders of magnitude smaller than the pressure gradients in the xylem of trees at maximum rates of transpiration, estimated to vary between 2 and $5\times10^4$ Pa.m$^{-1}$ (Nobel, 1999; Zimmermann and Brown, 1971).

Therefore, angiosperms that secrete kino, gum or gum resin holocrinously seemed to have found a way of defending themselves by accumulating and releasing their exudates under pressure within extensive vertical and tangential anastomosing networks, possibly with less metabolic energy costs. Unlike resin, gum, and gum resin ducts that are loaded granulocrinously, there is no energy spent on maintaining, through membrane channels or transporters, a pressure difference driving the flow between the surrounding epithelium and the kino vein lumen. The metabolic energy seems thus to be mostly spent on the meristematic activity of the epithelium and on the synthesis and accumulation of polyphenolic compounds in the vacuole and gummous compounds between the plasmalemma and the cell wall, which will be subsequently released holocrinously into the vein lumen after programmed cell autolysis. Combined, these processes seem to result in a minimal build-up of pressure necessary to drive flow through kino veins. Hence, relying on different duct development, dimensions, organization and distribution, the species that produce kino manage to deliver equally fast responses to damage as those observed on species producing resin granulocrinously. The range of values of the average flow velocity predicted by this model (Figs.

6A, 9) gives a mass flux between 0.1 and $4\times10^{-4}$ kg.s$^{-1}$.m$^{-2}$ for the dimensions of kino veins and pockets considered. This result agrees with experimental observations made on other species that also exude kino, e.g. *Eucalyptus sieberi* L.A.S.Johnson (Table 3). With slower flow rates, the metabolic energy costs will be even lower, possibly also translated into smaller rates of synthesis and accumulation of polyphenols, gummous components, or cell autolysis, as well as meristematic activity to produce smaller pressure gradients needed to drive flow holocrinously. This results in having smaller holocrine secretion rates, which is translated by having higher values of parameter K, which combined with wider veins confers them smaller specific resistance to flow (Fig. 4B), and consequently the need for smaller pressure gradients to drive flow through them. Judging from the few experimental observations on gum and kino flow rates, this seems to be the case in some species belonging to genera *Azadirachta*, *Butea* and *Moringa*, which are known to yield kino (Table 3).

## 3.2. Epithelial cells autolysis and the accumulation of polyphenolic and gummous components determine kino flow without changing its dynamics

The arrangement of anastomosing veins of different width, thus surrounded differently in terms of epithelial cell number, allows the same plant to exude varying amounts of kino through time depending on the number of different veins being ruptured. Larger veins surrounded by larger epithelia, i.e. presenting smaller surface-area-to-volume ratio, smaller values of the decay constant, $\frac{K\nu}{R^2}$, and higher values of the kino flow half-life, $t_{1/2}$, secrete more and for longer periods of time than narrower veins lined by smaller epithelia, thus presenting higher values of the decay constant and smaller values of the flow half-life. The decay constant is further decreased, that is, the flow half-life is increased, by smaller values of the kino kinematic viscosity, $\nu$, that is, for less viscous and denser kinos. Therefore, differences in the kino producing capacity (by volume and/or time) between species, or even individuals, could be then caused not only by different vein dimensions, but also differences in physiology through combinations of varying values of parameters K and *v*. This means that the different capacity to produce and exude kino holocrinously would ultimately reflect differences in anatomy and physiology, either through kino composition, which would be shown in its viscosity and is known to vary between species (Martius *et al*., 2012), or through the synthesis and accumulation of polyphenolic and gummous components, epithelium autolysis and meristematic activity or, most likely, a combination of all these factors, expressed in the values of parameter K. The values of the decay constant and kino flow half-life here obtained agree with the values estimated for these parameters from observations on other kino-producing species, e.g. *Acacia nilotica* (L.) Delile (Das, 2014). This result together with the estimates of mass flux obtained exemplifies how well the model here presented describes holocrine secretion and kino flow.

A 10 fold increase in the value of parameter K used in this model, i.e. up to $1\times10^{-6}$, meaning a higher decay constant reflecting a smaller holocrine loading rate, either due to smaller rates of synthesis and accumulation of polyphenolic and gummous components or epithelium autolysis or both, changes the kino vein specific resistance very little (Fig. 4B). Additionally, the increase in the vein specific resistance caused by decreasing the value of parameter K 10 times (Fig. 4B) does not change the order of magnitude of the pressure gradients needed to drive flow. Therefore, these results suggest that species that secrete kino holocrinously might react and respond differently to trauma by changing the rates of synthesis and accumulation of polyphenols and gummous components, thus affecting kino composition and consequently its viscosity, as well as epithelium autolysis, expressed in the values of parameter K, without changing the dynamics of flow. Thus, facing the likelihood of prolonged trauma by continuous stimuli, the possibility of having large kino pockets exuding continuously seems advantageous. This agrees with the variability of kino exudation observed on eucalypts (*Angophora*, *Corymbia*, and *Eucalyptus*), from small amounts being exuded in short periods of time to large volumes being exuded during longer periods (Hillis, 1987; Hillis and Hasegawa, 1963; Phillips, 1933). In this work, this is exemplified by comparing a narrower kino vein to a wider kino pocket that exudes for a much longer period of time (Fig. 4B). Due to their smaller specific resistance associated with higher holocrine secretion rates, pressure within wider veins and pockets decreases less with distance and increases more slowly with time than in narrower veins. This allows them to deliver equally fast responses as those of narrower veins, but with larger volumes under smaller pressure gradients. Ultimately, this results in having less energy costs needed to drive such flow through wider veins without changing its dynamics. Hence, the investment in maintaining wider ducts surrounded by larger epithelia that secrete holocrinously seems to compensate by keeping flow at lower costs. Therefore, compared to the narrower resin ducts in conifers that rely on granulocrine secretion, internal secretory systems that rely on holocrine secretion seem thus advantageous and may possibly constitute an evolutionary step of angiosperms in using internal secretory systems in plant defence mechanisms.

The overall duration of flow, $t_c$, up to 100 d in the kino pocket, and kino solidification rate, in the order of $10^{-11}$ $m^3.s^{-1}$, fall within the range of values obtained from the data of Das (2014) on *Acacia nilotica* (L.) Delile, a kino-producing species (Table 3). However, considering the differences in kino composition among species (Martius *et al.*, 2012) and the changes in the flow rate observed, e.g. on *Eucalyptus sieberi* L.A.S. Johnson (Hillis and Hasegawa, 1963), *Acacia nilotica* (L.) Delile (Das, 2014; Kuruwanshi, 2017)), *Acacia senegal* (L.) Willd. (Vasishth and Guleria, 2017), the solidification of kino may depend not only on its composition, but also on environmental factors, e.g. temperature, relative humidity, exposure to sunlight. Hence, the variability observed on the exudation of kino complicates its tapping and perhaps it is one of the reasons for not having more successful and sustainable methods developed so far. In conifers, resin flow rate is more or less uniform between the ruptured ducts. The amount of resin exuded, thus the magnitude and duration of the plant's response, depends on the species and duct density and width (Cabrita, 2018). In this respect, it would be helpful to investigate if and how the dimensions and distribution of kino veins relate to the type of stimulus, i.e. the nature of the trauma behind their origin, and consequently the ability of the plant to exude kino.

### 3.3. Kino role in plant defence mechanisms is enhanced by vein structure and organization

The chemical role played by kino on plant defence strategies, i.e. its biocidal properties against bacteria, fungi and virus (Locher and Currie, 2010; Martius *et al*., 2012), may be further enhanced physically by the contribution of kino vein structure and organization while acting as a barrier zone (Hillis, 1987; Tippett and Shigo, 1981). The outer layers of the peripheral cambium with highly suberized cells lining fully developed kino veins and pockets seem to behave like a periderm (Figs. 1D, 3). The main role of the periderm with its suberized cells is protective by preventing water and nutrient loss, as well as pathogen attack. In this regard, the periderm-like outer layer of cells lining fully developed kino veins seems to prevent not only water and solute loss from the surrounding axial parenchyma, but also possible incursions into it by pathogens, or substances released by them, which have successfully managed to penetrate deeper into kino veins. With highly suberized cells lining kino veins, water and solute exchange with the surrounding axial parenchyma and, subsequently, the rest of the plant is only possible through the symplast; thus becoming more limited. This gives the plant more control on the possible exchanges between the symplast and kino veins. Access to other regions of the plant body is further complicated by the sparse vascular system conducting elements present in the axial parenchyma surrounding kino veins. Considering the chemical properties of kino together with their distribution and structural organization, kino veins seem thus to form an effective barrier to harmful agents, especially against vascular diseases similarly to what is observed on other angiosperms with parenchymal barrier zones that do not produce kino, e.g. *Acer* sp., *Fagus sylvatica* L. (Torelli *et al*., 1994), *Ulmus americana* L., (Tippett and Shigo, 1981). In this respect, the presence of kino veins seems more advantageous compared to the apparent easier access offered by the resin ducts, lined by thin-walled epithelial cells densely connected to surrounding subsidiary cells that in turn connect to adjacent tracheids, i.e. vascular system (Ferreira and Tomazello-Filho, 2012).

## 4. Conclusions

Some angiosperms, here exemplified by species that produce kino veins as a response to trauma, seemed to have found an efficient way of using holocrine secretion as defence by combining physical and chemical strategies. Compared to conifers and other species that rely on pressure-driven loading of resin, gum or gum resin, these kino-producing species present an equivalent internal secretory system that delivers equally fast flows at lower pressure, driven by smaller pressure gradients, and possibly with lower energy costs to plant metabolism. Hence, with different organization, structure and physiology, kino veins may present a more efficient defence strategy of some angiosperms when compared to resin ducts. It would be then worthwhile to study and compare the effects of development, distribution, structure, and physiology of ducts on flow, and ultimately on plant defence, of other species that produce gum or gum resin. Understanding of how these physiological and morphological parameters affect gum flow might be useful to not only elucidate the plants abilities to resist different pathogens, but also select species or varieties, and develop more sustainable and economically viable methods of tapping gum and gum resin in angiosperms. Many of the tapping methods still in use today are not only unproductive, but also quite destructive and wasteful (Bhatt, 1987; Das, 2014; Nair, 2000, 2003).

# Appendix

From definition (4) and the governing Eq. (15) one finds that:

$$\frac{\partial^2 p}{\partial z \partial r} = 0 \tag{A1}$$

Differentiating governing Eq. (16) with respect to *r* and considering result (A1) one gets:

$$\frac{\partial^2 u_z}{\partial t \partial r} = \frac{\partial}{\partial r}\left[\frac{\nu}{r}\frac{\partial}{\partial r}\left(r\frac{\partial u_z}{\partial r}\right)\right] \tag{A2}$$

Let one assume that the axial velocity $u_z$ is of the form:

$$u_z(r,z,t) \equiv \phi(r)\gamma(z)\theta(t) \tag{A3}$$

Substituting it into equation (A2) one obtains:

$$\frac{d\phi}{dr}\frac{d\theta}{dt} = \theta\nu\frac{d}{dr}\left[\frac{1}{r}\frac{d}{dr}\left(r\frac{d\phi}{dr}\right)\right] \tag{A4}$$

which is a separable variable second order differential equation. Therefore, one has that:

$$\frac{d\theta}{dt} = -\kappa\nu\theta \tag{A5}$$

and

$$-\kappa \frac{d\phi}{dr} = \frac{d}{dr}\left[\frac{1}{r}\frac{d}{dr}\left(r\frac{d\phi}{dr}\right)\right] \tag{A6}$$

where $\kappa$ is a separation constant to be determined. Integration of equation (A5) over $t$ gives:

$$\theta(t) = \theta_0 e^{-\kappa \nu t} \tag{A7}$$

where $\theta_0$ is an undetermined constant. Integration of equation (A6) over $r$ gives:

$$\frac{d^2\phi}{dr^2} + \frac{1}{r}\frac{d\phi}{dr} + \kappa\phi = A \tag{A8}$$

where $A$ is a constant to be determined. The homogenous form of equation (A8) is a Bessel equation of order zero:

$$\frac{d^2\phi_H}{dr^2} + \frac{1}{r}\frac{d\phi_H}{dr} + \kappa\phi_H = 0 \tag{A9}$$

Considering that the axial velocity $u_z$ (A3) is bound at all points, including $r = 0$, the solution of the homogenous equation (A9) is:

$$\phi_H(r) = BJ_0\left(\sqrt{\kappa}\,r\right) \tag{A10}$$

where $B$ is a constant to be determined. Hence, the general solution of equation (A8) is of the form (Abramowitz and Stegun, 1972):

$$\phi(r) = \varsigma(r)\phi_H \tag{A11}$$

Substituting it into equation (A8) and considering the homogenous equation (A9) one obtains:

$$\frac{d}{dr}\left(r\phi_H^2 \frac{d\varsigma}{dr}\right) = Ar\phi_H \tag{A12}$$

Considering result (A10), the properties of Bessel functions (Abramowitz and Stegun, 1972) as wells as the no-slip boundary condition at the vein lumen boundary (5) (Fig. 3), $r = R$, (5), integration of equation (A12) over $r$ gives:

$$\phi(r) = DB\left(J_0\left(\sqrt{\kappa}\,r\right) - J_0\left(\sqrt{\kappa}\,R\right)\right) \tag{A13}$$

where $D$ is an undetermined constant. Therefore, considering results (A7) and (A13), the axial velocity $u_z$ (A3) is given by:

$$u_z(r,z,t) = DB\theta_0\left(J_0\left(\sqrt{\kappa}\,r\right) - J_0\left(\sqrt{\kappa}\,R\right)\right)\gamma(z)e^{-\kappa\nu t} \tag{A14}$$

Considering result (A7) and substituting $u_z$ (A14) and $u_r$ (4) into the continuity Eq. (2), one concludes that:

$$\gamma(z) = Ez + F \tag{A15}$$

where $E$ and $F$ are undetermined constants and

$$\frac{\partial}{\partial r}(r\xi) = EDB\left(rJ_0\left(\sqrt{\kappa}R\right) - rJ_0\left(\sqrt{\kappa}r\right)\right) \tag{A16}$$

Integration of equation (A16) over $r$ considering that the radial velocity $u_r$ is also bound at all points, including $r = 0$, gives:

$$\xi(r) = EDB\left(\frac{r}{2}J_0\left(\sqrt{\kappa}R\right) - \frac{J_1\left(\sqrt{\kappa}r\right)}{\sqrt{\kappa}}\right) \tag{A17}$$

Therefore, the axial velocity (A14) is written as:

$$u_z(r,z,t) = DB\theta_0\left(J_0\left(\sqrt{\kappa}r\right) - J_0\left(\sqrt{\kappa}R\right)\right)(Ez + F)e^{-\kappa vt} \tag{A18}$$

and the radial velocity (4) is given by:

$$u_r(r,z,t) = EDB\theta_0 \left( \frac{r}{2} J_0\left(\sqrt{\kappa}R\right) - \frac{J_1\left(\sqrt{\kappa}r\right)}{\sqrt{\kappa}} \right) e^{-\kappa vt} \tag{A19}$$

The average axial velocity is by definition given by:

$$\bar{u}_z(z,t) = \frac{2}{R}\int_0^R u_z(r,z,t)\, r\, dr = DB\theta_0 \left( \frac{2J_1\left(\sqrt{\kappa}R\right)}{\sqrt{\kappa}R} - J_0\left(\sqrt{\kappa}R\right) \right)(Ez+F)e^{-\kappa vt} \tag{A20}$$

Considering that the initial average axial flows at $z = 0$ and $z = L$ are known; i.e. $U_0$ (9) and $U$ (11) respectively, one finds that:

$$U_0 = DBF\theta_0 \left( \frac{2J_1\left(\sqrt{\kappa}R\right)}{\sqrt{\kappa}R} - J_0\left(\sqrt{\kappa}R\right) \right) \tag{A21}$$

and

$$U = DB\theta_0 \left( \frac{2J_1\left(\sqrt{\kappa}R\right)}{\sqrt{\kappa}R} - J_0\left(\sqrt{\kappa}R\right) \right)(EL+F) \tag{A22}$$

Combining both these results one obtains:

$$\frac{F}{E} = \frac{U_0 L}{U - U_0} \tag{A23}$$

Substituting this result into either results (A21) or (A22) one gets:

$$EDB\theta_0 = \frac{U - U_0}{L\left(\dfrac{2J_1(\sqrt{\kappa}R)}{\sqrt{\kappa}R} - J_0(\sqrt{\kappa}R)\right)} \tag{A24}$$

Therefore, considering results (A23) and (A24), the velocity components and the average axial velocity are given by respectively:

$$u_z(r,z,t) = \frac{(U-U_0)\left(J_0(\sqrt{\kappa}r) - J_0(\sqrt{\kappa}R)\right)}{L\left(\dfrac{2J_1(\sqrt{\kappa}R)}{\sqrt{\kappa}R} - J_0(\sqrt{\kappa}R)\right)} \left(z + \frac{U_0 L}{U - U_0}\right) e^{-\kappa \nu t} \tag{A25}$$

$$u_r(r,z,t) = \frac{(U-U_0)\left(\dfrac{r}{2}J_0(\sqrt{\kappa}R) - \dfrac{J_1(\sqrt{\kappa}r)}{\sqrt{\kappa}}\right)}{L\left(\dfrac{2J_1(\sqrt{\kappa}R)}{\sqrt{\kappa}R} - J_0(\sqrt{\kappa}R)\right)} e^{-\kappa \nu t} \tag{A26}$$

and

$$\bar{u}_z(z,t) = \left(\frac{U-U_0}{L}z + U_0\right)e^{-\kappa \nu t} \tag{A27}$$

While kino loading at the vein lumen boundary, $r = R$, is:

$$u_r(R,z,t) = -\frac{(U-U_0)R}{2L}e^{-\kappa vt} \qquad (A28)$$

Substituting the radial velocity (A26) into the governing Eq. (15) and integrating it over $r$ one finds that:

$$p(r,z,t) = \frac{(U-U_0)\kappa\mu J_0(\sqrt{\kappa}R)}{4L\left(\frac{2J_1(\sqrt{\kappa}R)}{\sqrt{\kappa}R} - J_0(\sqrt{\kappa}R)\right)} r^2 e^{-\kappa vt} + \omega(z,t) \qquad (A29)$$

where $\omega$ is an undetermined function. Substituting pressure (A29) and the axial velocity (A25) into the governing Eq. (16) and integrating it over $r$ gives:

$$\omega(z,t) = -\frac{(U-U_0)\kappa\mu J_0(\sqrt{\kappa}R)}{L\left(\frac{2J_1(\sqrt{\kappa}R)}{\sqrt{\kappa}R} - J_0(\sqrt{\kappa}R)\right)}\left(\frac{z^2}{2} + \frac{U_0 Lz}{U-U_0}\right)e^{-\kappa vt} + \lambda(t) \qquad (A30)$$

where $\lambda$ is an undetermined function of time. Pressure within the kino vein (A29) is thus given by:

$$p(r,z,t) = \frac{(U-U_0)\kappa\mu}{L\left(\frac{2J_1(\sqrt{\kappa}R)}{\sqrt{\kappa}R J_0(\sqrt{\kappa}R)} - 1\right)}\left(\frac{r^2}{4} - \frac{z^2}{2} - \frac{U_0 Lz}{U-U_0}\right)e^{-\kappa vt} + \lambda(t) \qquad (A31)$$

The average pressure is by definition:

$$\bar{p}(z,t) = \frac{2}{R}\int_0^R p(r,z,t)\,r\,dr = \frac{(U-U_0)\kappa\mu}{L\left(\dfrac{2J_1(\sqrt{\kappa}R)}{\sqrt{\kappa}RJ_0(\sqrt{\kappa}R)}-1\right)}\left(\frac{R^2}{8}-\frac{z^2}{2}-\frac{U_0Lz}{U-U_0}\right)e^{-\kappa\nu t}+\lambda(t)$$

(A32)

From boundary condition (10), at the origin, $z = 0$, one has that:

$$\lambda(t) = p_i - \frac{(U-U_0)\kappa\mu R^2}{8L\left(\dfrac{2J_1(\sqrt{\kappa}R)}{\sqrt{\kappa}RJ_0(\sqrt{\kappa}R)}-1\right)}e^{-\kappa\nu t} \qquad (A33)$$

Therefore, pressure (A31) and the average pressure (A32) within the kino vein can be written respectively as:

$$p(r,z,t) = p_i + \frac{(U-U_0)\kappa\mu}{L\left(\dfrac{2J_1(\sqrt{\kappa}R)}{\sqrt{\kappa}RJ_0(\sqrt{\kappa}R)}-1\right)}\left[\frac{2r^2-R^2}{8}-\frac{(U-U_0)z^2+2U_0Lz}{2(U-U_0)}\right]e^{-\kappa\nu t} \quad (A34)$$

and

$$\bar{p}(z,t) = p_i - \frac{\kappa\mu\left[(U-U_0)z^2 + 2U_0 Lz\right]}{2L\left(\dfrac{2J_1(\sqrt{\kappa}R)}{\sqrt{\kappa}R J_0(\sqrt{\kappa}R)} - 1\right)} e^{-\kappa\nu t} \tag{A35}$$

Considering the nature of holocrine secretion, one has that wider kino veins surrounded by a larger epithelial area present higher loading rates (A28) than those of narrower veins. That is, for two veins of radii $R_1$ and $R_2$, with $R_1 > R_2$, one finds that:

$$u_{r1}(R_1, z, t) > u_{r2}(R_2, z, t) \tag{A36}$$

hence

$$\kappa_2 > \kappa_1 \tag{A37}$$

Therefore, for two kino veins of equal length $L$ but different radii, $R_1 > R_2$, running close to each other and submitted to the same average pressure gradient, one has from equations (A27) and (A35) that:

$$\frac{\kappa_1}{\left(\dfrac{2J_1(\sqrt{\kappa_1}R_1)}{\sqrt{\kappa_1}R_1 J_0(\sqrt{\kappa_1}R_1)} - 1\right)} \bar{u}_{z1}(L,t) = \frac{\kappa_2}{\left(\dfrac{2J_1(\sqrt{\kappa_2}R_2)}{\sqrt{\kappa_2}R_2 J_0(\sqrt{\kappa_2}R_2)} - 1\right)} \bar{u}_{z2}(L,t) \tag{A38}$$

Considering result (A36) and its implications on the volume delivered by both veins at the wound, $z = L$, i.e.:

$$\bar{u}_{z1}(L,t) > \bar{u}_{z2}(L,t) \tag{A39}$$

one concludes from results (A37), (A38), and (A39) that:

$$\frac{2J_1\left(\sqrt{\kappa_1}R_1\right)}{\sqrt{\kappa_1}R_1 J_0\left(\sqrt{\kappa_1}R_1\right)} < \frac{2J_1\left(\sqrt{\kappa_2}R_2\right)}{\sqrt{\kappa_2}R_2 J_0\left(\sqrt{\kappa_2}R_2\right)} \tag{A40}$$

That is:

$$\sqrt{\kappa_1}R_1 < \sqrt{\kappa_2}R_2 \tag{A41}$$

This result suggests that parameter $\kappa$ relates inversely with the square of the radius, $R^2$, which in its simpler form can be considered as:

$$\kappa \equiv \frac{K}{R^2} \tag{A42}$$

where K is an undetermined constant. Using definition (A42), the velocity components (A25) and (A26), the average axial velocity (A27) and kino loading (A28) can be written, respectively, as:

$$u_z(r,z,t) = \frac{(U-U_0)\left(J_0\left(\frac{\sqrt{K}}{R}r\right) - J_0\left(\sqrt{K}\right)\right)}{L\left(\frac{2}{\sqrt{K}}J_1\left(\sqrt{K}\right) - J_0\left(\sqrt{K}\right)\right)}\left(z + \frac{U_0 L}{U-U_0}\right)e^{-\frac{K\nu}{R^2}t} \qquad (A43)$$

$$u_r(r,z,t) = \frac{(U-U_0)\left(\frac{r}{2}J_0\left(\sqrt{K}\right) - \frac{R}{\sqrt{K}}J_1\left(\frac{\sqrt{K}}{R}r\right)\right)}{L\left(\frac{2}{\sqrt{K}}J_1\left(\sqrt{K}\right) - J_0\left(\sqrt{K}\right)\right)}e^{-\frac{K\nu}{R^2}t} \qquad (A44)$$

$$\bar{u}_z(z,t) = \left(\frac{U-U_0}{L}z + U_0\right)e^{-\frac{K\nu}{R^2}t} \qquad (A45)$$

and

$$u_r(R,z,t) = -\frac{(U-U_0)R}{2L}e^{-\frac{K\nu}{R^2}t} \qquad (A46)$$

with pressure (A34) and the average pressure (A35) being given by:

$$p(r,z,t) = p_i + \frac{(U-U_0)\mu K}{LR^2\left(\frac{2J_1\left(\sqrt{K}\right)}{\sqrt{K}J_0\left(\sqrt{K}\right)} - 1\right)}\left[\frac{2r^2 - R^2}{8} - \frac{(U-U_0)z^2 + 2U_0 Lz}{2(U-U_0)}\right]e^{-\frac{K\nu}{R^2}t} \qquad (A47)$$

and

$$\overline{p}(z,t) = p_i - \frac{\mu K \left[(U - U_0)z^2 + 2U_0 L z\right]}{2LR^2 \left(\dfrac{2J_1\sqrt{K}}{\sqrt{K}J_0\sqrt{K}} - 1\right)} e^{-\frac{Kv}{R^2}t} \qquad (A48)$$

**Note**: Figures 1D, 1E, and 7 should be in colour online-only.

**Fig. 1** – Transverse view of the development of kino veins: (A) rosette, composed of darkly stained developing traumatic parenchyma cells with a dense cytoplasm and accumulating polyphenolic compounds (arrows), and lysis of the central cells of the rosette with the consequent formation of the lacuna (arrowhead) in the secondary xylem of *Azadirachta indica* A.Juss., very close to cambial zone; (B) developing kino vein lumen surrounded by darkly stained epithelial cells in *Azadirachta indica* A.Juss. (arrowhead); (C) fully developed kino veins in the secondary xylem of *Azadirachta indica* A.Juss., with lumina filled with phenolic compounds and loosely arranged cells (arrows) surrounded by lignified cells; (D) suberized cells (arrow) lining mature kino veins of *Eucalyptus obliqua* L'Hér.; (E) mature kino veins in the xylem of *Eucalyptus obliqua* L'Hér. Scale bar = 75 μm in A, B, and C, and scale bar = 200 μm in D and E. *C* = lacuna, *CZ* = cambial zone, *DX* = differentiating xylem, *KL* = kino vein lumen, *PB* = parenchyma bridge, *X* = xylem. (A) and (B) reprinted and figure (C) adapted from Journal of Sustainable Forestry, 20/2, K. S. Rajput, K. S. Rao, H. P. Vyas, Formation of gum ducts in *Azadirachta indica* A.Juss, pp. 1-13, Copyright (2005), with permission from Taylor & Francis Ltd, http://www.tandfonline.com. (D) and (C) adapted from IAWA Journal, 23/4, A. Eyles, C. Mohammed, Comparison of cepa (2-chloroethyl phosphonic acid) induced responses in juvenile *Eucalyptus nitens*, *E. globulus* and *E. obliqua*: a histochemical and anatomical study, pp. 419-430, Copyright (2002), with permission from Brill Academic Publishers, http://www.brill.com.

**Fig. 2** – Distribution of mature kino veins in the stem of a specimen of *Eucalyptus sieberi* L.A.S.Johnson injured in frequent fires (scale bar = 5 cm). Inset shows in the same specimen a set of kino veins, of which some are fused and enlarged into a kino pocket (scale bar = 1 cm). Reprinted by permission from Springer Nature Customer Service Centre GmbH: Springer-Verlag Berlin

Heidelberg, Heartwood and Tree Exudates by W. E. Hillis, Copyright (2003), https://doi.org/10.1007/978-3-642-72534-0.

**Fig. 3** – Model of a kino vein cross section of radius *R*.

**Fig. 4** – (A) Ratio between the kino vein specific resistance, $\sigma$, and the specific resistance of a similar duct under Poiseuille regime, $\sigma_{Poiseuille}$, for a range of values of parameter K. (B) Kino vein specific resistance, $\sigma$, for the range of values of the radius *R* observed on eucalypts and specific values of parameter K.

**Fig. 5** – Average pressure $\bar{p}$ (A) and pressure gradient (B) at various times in a 0.5 cm radius and 1 m length kino vein filled with kino of dynamic viscosity $\mu = 10$ Pa.s, kinematic viscosity $v = 7.14 \times 10^{-3}$ m$^2$.s$^{-1}$, ruptured at the end of its length, at $t = 0$ h, where kino is exposed to air at constant normal atmospheric pressure $p_a = 1$ atm, considering: $K = 1.0 \times 10^{-7}$; $U = 1.0 \times 10^{-4}$ m.s$^{-1}$; and $U_0 = 1.0 \times 10^{-6}$ m.s$^{-1}$.

**Fig. 6** – Average axial velocity $\bar{u}_z$ (A) and kino loading (B) in the kino vein described in Fig. 5 and in a 3.0 cm radius and 1 m length kino pocket under the same conditions as the kino vein.

**Fig. 7** – Axial velocity $u_z$ profile 1 h (black lines) and 8 h (red lines) after rupturing in the kino vein described in Fig. 5 at different points expressed as fractions of its total length $L = 1$ m.

**Fig. 8** – Average pressure $\bar{p}$ (A) and pressure gradient (B) at various times in the kino pocket described in Fig. 6.

**Fig. 9** – Average axial velocity $\bar{u}_z$ in the kino pocket described in Fig. 6.